\documentclass[10pt]{article} 

\usepackage[titletoc, title]{appendix}
\usepackage{titletoc}

\usepackage[utf8]{inputenc} 
\usepackage{setspace}

\usepackage{geometry} 
\usepackage{graphicx} 
 \usepackage[parfill]{parskip} 
 
\usepackage{booktabs} 
\usepackage{array} 
\usepackage{paralist} 
\usepackage{verbatim} 

\usepackage{yfonts}
\usepackage{hyperref}
\usepackage{caption}
\usepackage{subcaption}
\usepackage[affil-it]{authblk}

\usepackage{fancyhdr} 
\usepackage{sectsty}
\usepackage{amsmath}
\usepackage[makeroom]{cancel}
\allsectionsfont{\sffamily\mdseries\upshape} 
\usepackage{dsfont}
\usepackage{amssymb}
 \usepackage{epstopdf}
 \usepackage{minitoc}
     \usepackage{setspace}

\usepackage[nottoc,notlof,notlot]{tocbibind} 
\usepackage[titles,subfigure]{tocloft} 
\usepackage{graphicx} 
\usepackage{color}

\newcommand{\angler}{\rangle}
\newcommand{\anglel}{\langle}
\newcommand{\rrvert}{\vert}
\newcommand{\rrVert}{\Vert}
\newcommand{\llvert}{\vert}
\newcommand{\llVert}{\Vert}

\newcommand{\tr}{\operatorname{tr}}
\newcommand{\In}{\operatorname{In}}

\title{Trapped 
Modes in  Linear Quantum Stochastic Networks with Delays}
\author{Gil Tabak%
  \thanks{Electronic address: \texttt{tabakg@stanford.edu}}
  \vspace{-2ex}}
         \author{Hideo Mabuchi%
  \thanks{Electronic address: \texttt{hmabuchi@stanford.edu}}}
\affil{
Edward L.\ Ginzton Laboratory and Department of Applied Physics\\ Stanford University}

\begin{document}
\maketitle

\ifdefined\TOC
\begingroup
\def\addvspace#1{}
\tableofcontents
\endgroup
\fi

\begin{abstract}

Networks of open quantum systems with feedback have become an active
area of research for applications such as quantum control, quantum
communication and coherent information processing. A canonical
formalism for the interconnection of open quantum systems using quantum
stochastic differential equations (QSDEs) has been developed by Gough,
James and co-workers and has been used to develop practical modeling
approaches for complex quantum optical, microwave and optomechanical
circuits/networks.
In this paper we fill a significant gap in existing methodology by
showing how trapped modes resulting from feedback via coupled channels
with finite propagation delays can be identified systematically in a
given passive linear network. Our method is based on the
Blaschke-Potapov multiplicative factorization theorem for inner
matrix-valued functions, which has been applied in the past to analog
electronic networks. Our results provide a basis for extending the
Quantum Hardware Description Language (QHDL) framework for automated
quantum network model
construction (Tezak \textit{et al.} in Philos. Trans. R. Soc. A, Math. Phys. Eng. Sci.
370(1979):5270-5290, \cite{Tezak2012}
to
efficiently treat scenarios in which each interconnection of components
has an associated signal propagation time delay.


\end{abstract}

\section{Introduction}

\ifdefined\DEBUG
\else
\fi

%
%
%

Just as in classical electrical and light-wave circuit design, there
are many quantum network modeling scenarios in which it is necessary to
capture the impact of time delays in the propagation of signals between
components. For example in large-area communication networks there is
an obvious need to analyze synchronization issues; in integrated
photonic circuits the high natural bandwidth of nanoscale components
may create problems of delay-induced feedback instability, and may
support the design of devices (such as oscillators) that exploit finite
optical propagation delays.

In considering how best to represent and simulate time delays in
quantum networks we would like to strike an expedient balance between
the need to minimize additional computational overhead and the desire
to derive intuitive approximate models. Our main interest in this paper
is to develop a systematic approach to modeling the leading-order
effects of signal propagation time delays in a linear passive
quantum optical network that can be specified naturally using the
so-called SLH formalism of Gough, James and co-workers  \cite
{Gough2009a,Gough2009,Gough2008,Gough2010,Yanagisawa2003,Zhang2012}.
Whereas series and feedback interconnections of open quantum systems in
the SLH formalism are generally treated as having vanishing signal
propagation time delay, we seek to expand the formalism in a natural
way that allows each interconnection to have an associated finite delay.
Our approach utilizes additional degrees of freedom to capture the
behavior of trapped resonant modes created by the system's internal
network of feedback pathways and time delays, targeting a specific
frequency range that corresponds to the intrinsic bandwidths of
components in the network.

The study of time-delay systems has a long history (see \cite
{Richard2003} for a thorough overview).
In the context of quantum systems, a quantum control scenario
incorporating time-delayed coherent feedback has been analyzed recently
by Grimsmo \cite{Grimsmo2015}. The construction developed by
Grimsmo appears most suitable for use in scenarios with very large
feedback time delay, requiring a much higher computational overhead
than should be necessary when propagation delays are relatively small.
Very recently Pichler and Zoller \cite{PichlerZoller2015} have
described an approach to modeling the dynamics of a finite-delay
quantum channel that exploits the Matrix Product State formalism for
computational efficiency, and demonstrate its use in analyzing quantum
feed-forward and feedback dynamics. Our work here is distinguished by
showing how networks incorporating feedback via many coupled signal
channels can be treated efficiently and by focusing on SLH-compatible
modeling at the level of quantum stochastic differential equations
(QSDEs). The method we present can straightforwardly be incorporated
into the Quantum Hardware Description Language (QHDL) framework \cite
{Tezak2012} for automated model construction for complex quantum networks.

\section{Preliminaries}
In the spirit of SLH/QHDL we assume that we are given a network of open
quantum systems whose
input ports, output ports, and static passive linear components are
connected over some channels representing the signal propagating in the
network.
We additionally assume that an end-to-end propagation time delay is
specified for each channel. If feedback loops are created by the
network topology, trapped modes will be created that may need to be
modeled dynamically in order to accurately simulate the overall
behavior of the network.

The basic problem lies in choosing a procedure for embedding new
stateful dynamics into the `space between components' in an SLH
network. Prior work such as \cite{PichlerZoller2015} has addressed the
question of how to model an individual channel with finite time delay
efficiently, however, our philosophy here will be to work at the level
of more complex sub-networks that mediate interconnections among
multiple-input/multiple-output components. Our method is restricted to
sub-networks that are linear and passive, and thus may include
components such as beam-splitters and phase shifters but not, \textit{e.g.}, gain elements or nonlinear traveling-wave interactions. We
nevertheless gain a significant advantage by considering linear passive
sub-networks in that we are able to recognize the creation of trapped
modes by feedback with finite time delays, and can provide a systematic
procedure for adding the stateful dynamics required to simulate the
behavior of such modes within a frequency band of interest.

We treat channels as passive linear quantum stochastic systems \cite
{Maalouf2009,Maalouf2011,Petersen2010,Petersen2014} whose input-output
behavior can be characterized by the relationship of the input and
output annihilation fields only. The input-output relationship for a
linear system must satisfy certain physical realizability conditions.
Our systematic method preserves the physical realizability condition
while allowing us to simulate the system with only a small number of
degrees of freedom.
Our resulting approximate system describing the dynamics of a
passive linear sub-network can also be combined with other possibly
nonlinear components using the standard SLH composition rules.
Incorporating nonlinear components embedded within a network whose
topology results in trapped modes may need a more delicate treatment.
To do this, the interaction Hamiltonian between the trapped modes and
the nonlinear components must be found. This is an issue we will write
about in more detail in a future publication.


Below, the system studied is a particular sub-network of the kind
we discussed above.
Essentially our approach introduces an approximation with finitely many
state-space variables to a given system (we will refer to such a system
as a finite-dimensional system).
For a system with $N$ input and output ports and $M$ oscillator modes
$a_1,\ldots,a_M$ satisfying the canonical commutation relations
%
\begin{align}
\bigl[a_i,a_j^*\bigr] = \delta_{ij},\qquad
[a_i,a_j] = 0, \qquad \bigl[a_i^*,a_j^*
\bigr] = 0,
\end{align}
a passive linear model can be described by the input-output relation
%
\begin{align}
\label{equ:input_output} %
\begin{pmatrix}
d\mathbf{a}(t) \\
d \mathbf{B}_{\mathrm{out}}(t)
\end{pmatrix} %
= %
\begin{pmatrix}
A & B \\ C & D
\end{pmatrix} %
\begin{pmatrix}
\mathbf{a}(t)\,dt\\
d \mathbf{B}(t)
\end{pmatrix} .
\end{align}
The bold font used here denotes vectors.
The term
$\mathbf{a}(t)$ above represents the oscillator modes in the Heisenberg picture.
The $A$, $B$, $C$, $D$ are complex-valued matrices of appropriate size.
The $B_i$ and $B_{\mathrm{out},i}$ are the forward differentials of adapted
quantum stochastic processes satisfying the commutation relations with their
respective operator adjoints, in the sense of \cite
{Hudson1984,Parthasarathy1992}:
%
\begin{align}
\bigl[b_i(t),b_j^*(s)\bigr] = \delta(s-t)
\delta_{ij},\qquad \bigl[b_i(t),b_j(s)\bigr] =
0, \qquad \bigl[b_i^*(t),b_j^*(s)\bigr] = 0,
\end{align}
where formally $\mathbf{b}(t) = d\mathbf{B}(t)/{dt}$ is the white noise
operator.
We will refer to this formulation as the state-space representation, or
the $ABCD$ formulation, of the system.

The matrices in the $ABCD$ formulation here are related to the SLH
model by
%
\begin{align}
S =D, \qquad L = C \mathbf{a}, \qquad H = \mathbf{a}^{\dagger}\Omega
\mathbf{a},
\end{align}
where
%
\begin{align}
A = -\frac{1}{2} C^{\dagger} C - i\Omega, \qquad B =
-C^{\dagger} S.
\end{align}
Here, $\Omega$ is a Hermitian matrix.
An introduction for passive linear systems can be found for example in
\cite{Guta2014}.

Our approach seeks to approximate the transfer function within a given
frequency range by selecting only a finite subset of the original modes
and generating a state-space representation using the information close
to the zeros or poles of the modes.
The resulting approximation is a passive linear system satisfying the
physical realizability condition.
We discuss a sufficient condition for our approximation to converge to
the true transfer function for a large class of possible transfer functions.

There are other approaches that could be used to obtain a different
set of modes that may approximate the system of interest. For example,
one approach may involve approximating each delay term in the system
using a symmetric Pad\'{e} approximation, which would result in a
physically realizable component (see Appendix~\ref{section:pade_approximation}). Although this approach can be simple to
use, the Pad\'{e} approximation will not always introduce the zeros and
poles of the transfer function at the correct locations, and may
introduce spurious zeros and poles to the approximated transfer
function. On the other hand, the zero-pole interpolation by
construction adds zeros and poles to the approximated transfer function
only when they are present in the original transfer function. This
feature of our approach may be important in many physical applications
because the locations of the zeros and poles have physically meaningful
consequences, including the resonant frequencies and linewidths of the
effective trapped cavity modes resulting in the network due to
feedback. Nevertheless, as discussed in Section~\ref{section:singular},
there are passive linear systems for which the zero-pole interpolation
is insufficient - in this case, a finite-dimensional state-space
representation will necessarily have spurious zeros and poles.

\section{Problem Characterization}
\label{section:problem_characterization}
\subsection{Frequency domain}
Throughout we work in the frequency domain (specifically in the
$s$-domain unless otherwise noted).
A function of time $f(t)$ can be mapped to the frequency domain by the
Laplace transform, resulting in a function $F(z)$. The Laplace
transform is given by
%
\begin{align}
F(z) =
\int_{-\infty}^\infty e^{-zt} f(t)\,dt.
\end{align}
Here, $z = \sigma+ i \omega$ for $\sigma,\omega\in\mathbb{R}$.
When $\sigma= 0$, $z = i \omega\in i \mathbb{R}$ represents a real frequency.
The input-output relation of a linear system can be characterized in
the frequency domain.
This relation between inputs and outputs in the frequency domain is
captured by the \emph{transfer function} $T(z)$. The transfer function
is defined by the relation between the inputs $\mathbf{I(z)}$ and
outputs $\mathbf{O(z)}$ by $\mathbf{O}(z) = T(z) \mathbf{I}(z)$.
For example,
we can find the transfer function of the system described by
Eq. (\ref{equ:input_output}) by taking the Laplace transform of the
equation. After some algebra, the transfer function is found to be
%
\begin{align}
T(z) = C(zI-A)^{-1}+D.
\end{align}

\subsection{Problem characterization in the frequency domain}
We consider a linear system with $N$ input and $N$ output ports. We
will primarily be interested in a system which is linear, passive, and
has a transfer function
$T(z)$ that is unitary for all $z\in i \mathbb{R}$. The last condition
guarantees that the system conserves energy. We remark that more
generally the loss of energy can be considered for example by adding
additional ports.
We assume throughout the transfer function is a meromorphic
matrix-valued function. For simplicity, we assume that each pole has
multiplicity one.


We will be particularly interested in a system consisting of time
delays and beamsplitters. A delay of length $T$ has transfer function
of the form $e^{-zT}$.
In general, any system of delays and beamsplitters can be written as
%
\begin{align}
\label{equ:delays_and_beamsplitter} %
 \begin{pmatrix}
x \\
x_{\mathrm{out}}
\end{pmatrix} %
= %
\begin{pmatrix}
M_1 & M_2 \\
M_3 & M_4
\end{pmatrix} %
\begin{pmatrix}
E(z) x \\
x_{\mathrm{in}}
\end{pmatrix} %
.
\end{align}

\begin{figure}
    \centering
\includegraphics[width=0.4\textwidth]{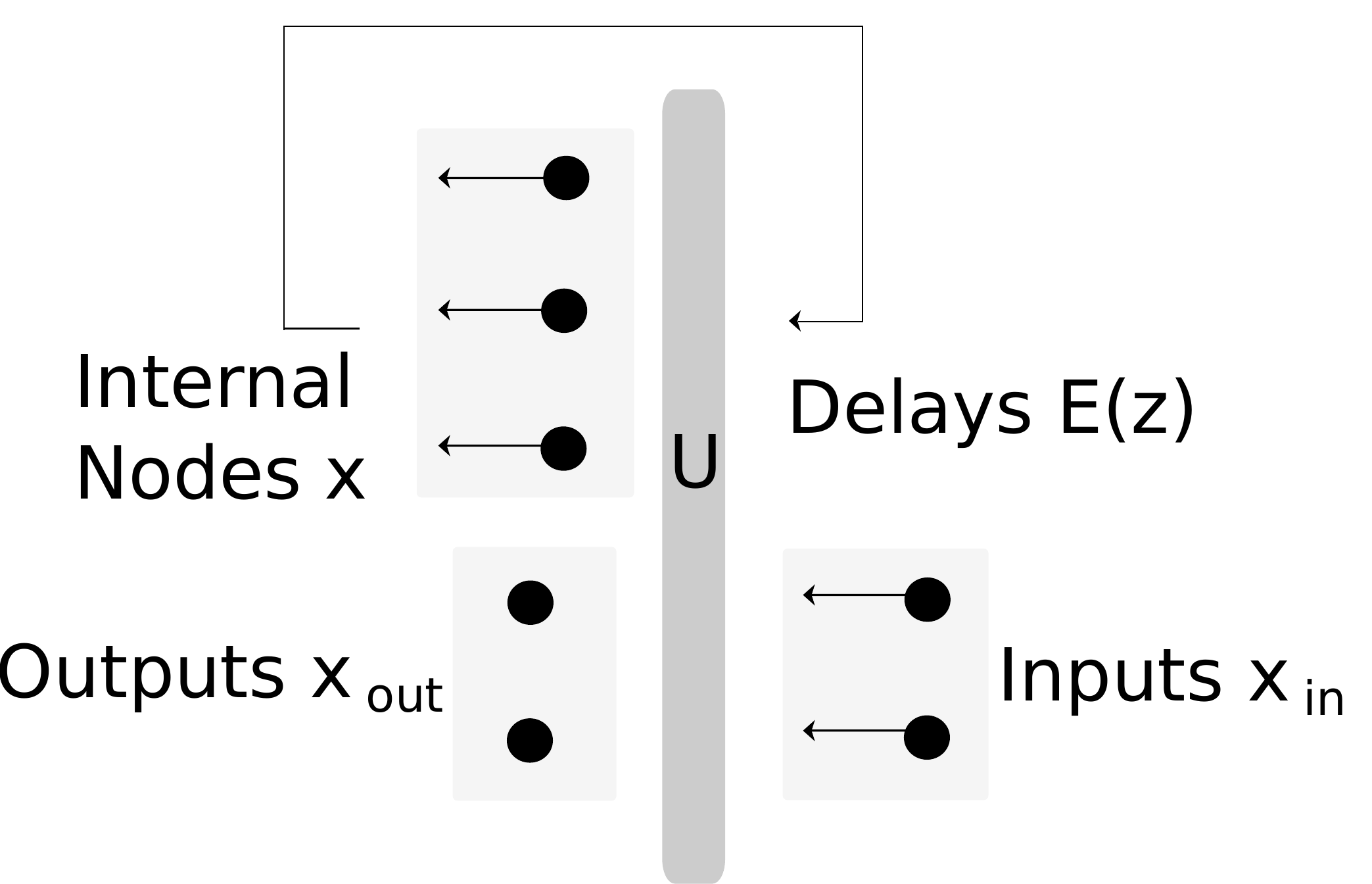}
\caption{\textbf{Schematic setup of the system.} This network is described by
Eq. (\ref{equ:passive_TF}). The $U$ in the figure represents a unitary component.}\label{fig:set-up}
\end{figure}

Here $x_{\mathrm{out}}$ and $x_{\mathrm{in}}$ are respectively the
$N$-dimensional input and output signals, and $x$ are the internal
signals of the system. The values of $x$ are taken along edges
corresponding to delays before each signal is delayed.  The $M_i$ are
constant matrices of the appropriate size determined by the specific
details of the system. $E(z)$ is a diagonal matrix whose diagonal
elements are the transfer functions of the various delays in the
system, $e^{-zT_1},e^{-zT_2},\ldots,e^{-zT_N}$. The system is illustrated
abstractly in Figure~\ref{fig:set-up}.

The transfer function of the given system can be formally solved as
%
\begin{align}
\label{equ:passive_TF} x_{\mathrm{out}} &= T(z) x_{\mathrm{in}} =
\bigl[M_3 E(z) \bigl(I-M_1 E(z)\bigr)^{-1}
M_2 +M_4\bigr] x_{\mathrm{in}}
\\
&= \bigl[M_3 \bigl(E(-z)-M_1 \bigr)^{-1}
M_2 +M_4\bigr] x_{\mathrm{in}}.
\end{align}
Notice that this transfer function will have poles when
%
\begin{align}
\label{equ:poles} \det \bigl(I-M_1 E(z) \bigr) = 0.
\end{align}
We can define the zeros of this transfer function as $z$ satisfying
$\det (T(z)) = 0$.

Throughout, we will also assume that the system is asymptotically
stable. In terms of the network transfer function, all the eigenvalues
of $M_1$ have norm less than $1$, with the consequence that $T(z)$ is
bounded for $\Re(z) \ge0$.

This transfer function has the important feature of being unitary
whenever $z$ is purely imaginary. That is,
$T(i\omega)T^{\dagger}(i\omega) = T^{\dagger}(i\omega)T(i\omega)= I$ whenever
$\omega\in\mathbb{R}$.
This is exactly the physical realizability condition for a passive
linear system.
We will refer to this constraint throughout the paper as the \emph
{unitarity constraint}.
One consequence of this condition that can be obtained by analytic
continuation is that
$T(z) T^{\dagger}(-\overline{z}) = I$ except when a pole is encountered.

\emph{Some observations.}
We see that the poles and zeros occur in pairs $z$, $-\overline{z}$.
In this paper we will refer to such a pair as a \emph{zero-pole pair}.
In general, we observe that there may be infinitely many solutions to
Eq. (\ref{equ:poles}). If we take $M_1$ to be a real matrix, $z$ is a
pole whenever $\overline{z}$ is also a pole. Furthermore, applying the
maximum modulus principle shows the system is stable (see Appendix~\ref{section:max}). This implies that the poles appear in the left half-plane.

The solution to Eq. (\ref{equ:poles}) can be found numerically within a
bounded subset of $\mathbb{C}$. There are dedicated algorithms that use
contour integration to guarantee finding all the roots within a
contour, which we briefly discuss in Section~\ref{section:contour}.
In the special case when the time delays are commensurate, we can
re-write the equation for the poles as a polynomial equation of the
variable $w = e^{-zT_0}$ for some $T_0$. Doing this will make the root
finding procedure much more simple.

If the system has passive linear components other than delays and
beamsplitters, the transfer function shown above may be modified. It
will still have the same important features that the poles appear on
the left half-plane (except when the system is marginally stable) and
the function restricted to the imaginary axis is unitary.

\section{Specific example and approximation procedure: one time delay
and one beamsplitter}
\label{section:first_example}

\begin{figure}[b]
    \centering
    \includegraphics[width=0.4\textwidth]{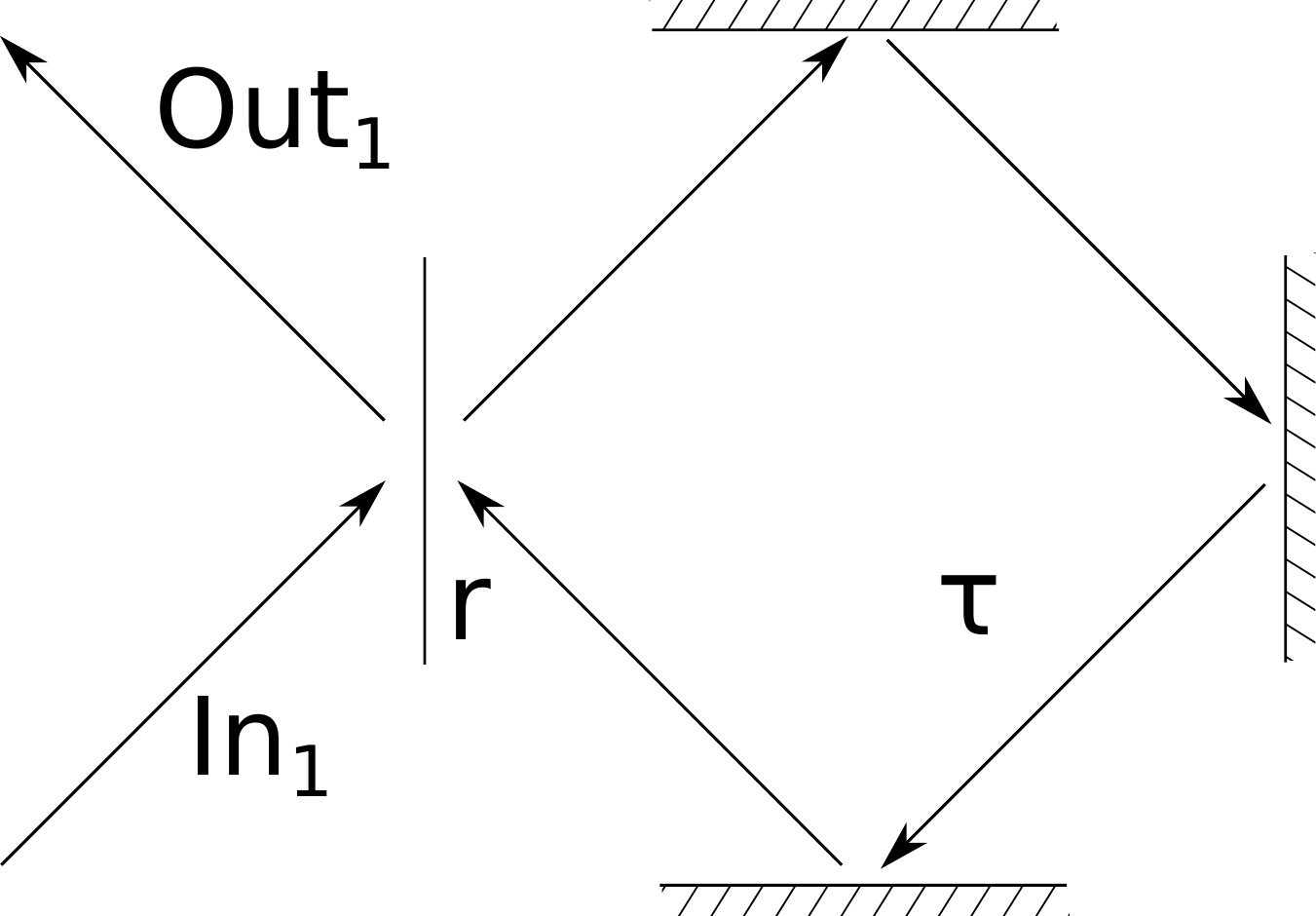}
\caption{\textbf{A cavity  formed by a single time delay $\tau$ and a single beamsplitter
with reflectiveness $r$.}}\label{fig:Time_Delay}
\end{figure}

We will consider the example illustrated in Figure~\ref{fig:Time_Delay}, where a single beamsplitter and time delay are
combined to form a single-input and single-output (SISO) system.
The example here is analogous to the one considered in Section VII B of
\cite{Gough2010}.

\begin{figure}
\centering
    \includegraphics[width=0.6\textwidth]{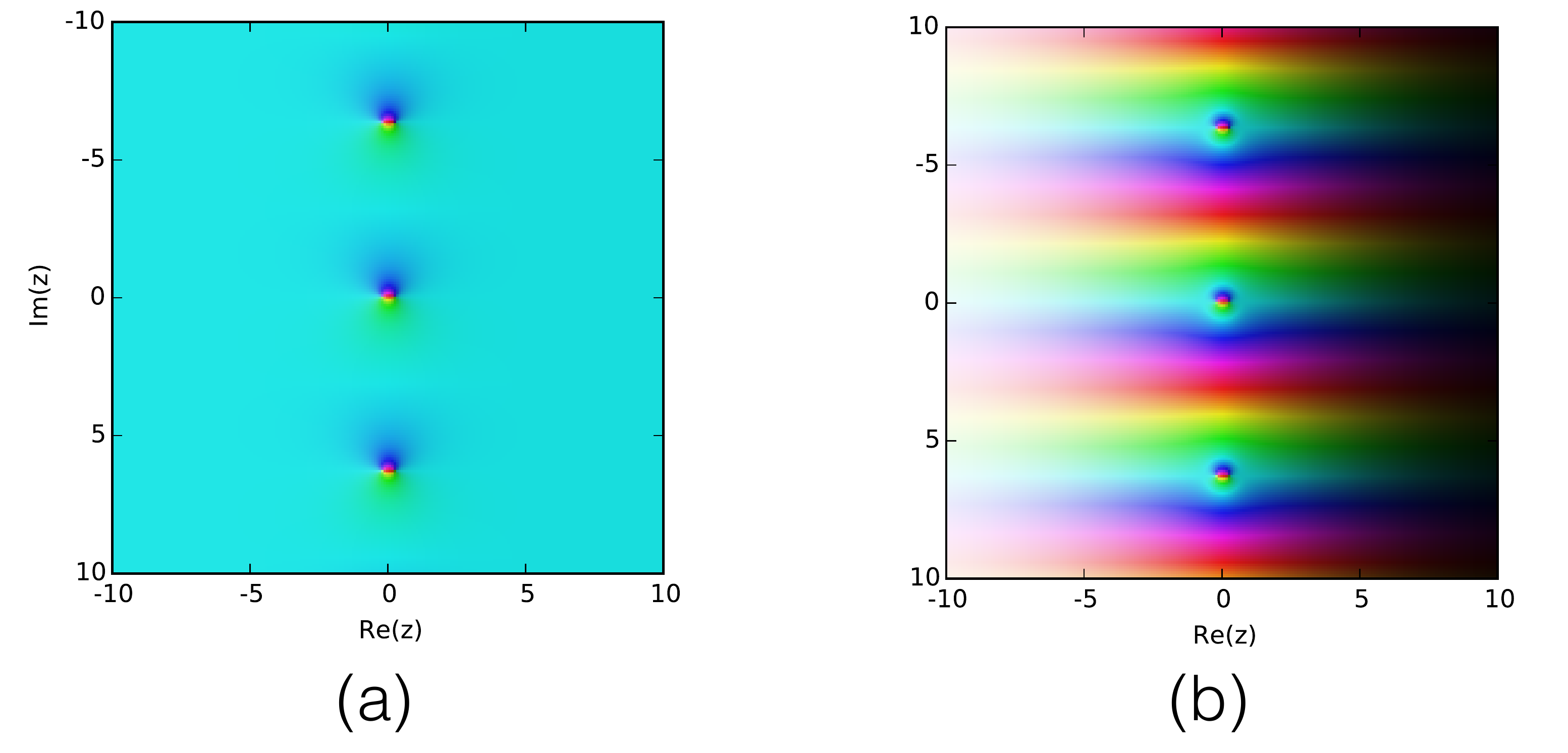}
\caption{\textbf{A comparison of two transfer functions as seen in the complex plane.} The hue and
the brightness correspond to the phase and magnitude, respectively. \textbf{(a)} The transfer function
$T(z) = \frac{ e^{- z} - r}{1-r e^{- z}}$  corresponds to a cavity with $r = 0.8$ and $\tau = 1$.
\textbf{(b)} The transfer function $T(z) = \frac{ e^{- z} - r}{1-r e^{- z}} e^{-z}$ in addition  to a cavity augments
a delay of length $\tau$ in sequence with the original cavity system.}\label{fig:eg_complex}
\end{figure}

In this particular example, we use the convention that a
beamsplitter has the transfer function
\begin{align*}
T_{BS} = %
\begin{pmatrix}
t & -r \\
r & t
\end{pmatrix}, %
\end{align*}
where $r^2 + t^2 =1$,
and a time delay of length $\tau$ has the transfer function
\begin{align*}
T_\tau(z) = e^{-\tau z}.
\end{align*}
The transfer function of the system drawn in Figure~\ref{fig:Time_Delay} is given by
%
\begin{align}
\label{equ:cavity} T(z) = \frac{ e^{-\tau z} - r}{1-r e^{-\tau z}}.
\end{align}
This transfer function is illustrated in Figure~\ref{fig:eg_complex}(a).
The poles $p_n$ for $n \in\mathbb{Z}$ are found to be
%
\begin{align}
p_n = \frac{1}{\tau} \bigl( \ln(r) + 2 \pi i n \bigr).
\end{align}
Notice that the real part is negative, as it should be for a stable system.
Each of the poles in the system here corresponds to a cavity mode. The
imaginary part corresponds to the mode frequency and the real part
determines the linewidth properties. With this interpretation, we can
readily relate the free spectral range to the delay length. In an
attempt to approximate the system, we can consider a product of the
following form of terms having the same (simple) poles as $T(z)$,
satisfying the unitarity condition, and agreeing in value with $T(z)$
at $z = 0$
%
\begin{align}
\label{equ:prod_ex} 
\tilde T(z) =-\prod
_{n \in\mathbb{Z}}c_n \biggl( \frac{z+\overline{p_n}
}{z-p_n }
\biggr).
\end{align}
The factors $c_n$ are phases to ensure the convergence of the product.
If we evaluate the product as a limit of products $\tilde T_N(z)$ as
$N\to\infty$ such that for a fixed $N$, $\tilde T_N(z)$ is a product
over $n=-N,-N+1,\ldots,N-1,N$, we can drop the $c_n$ factors by symmetry
considerations.

One interpretation of each term in the product is the frequency-domain
representation of a cavity mode. We can obtain an approximation for
$T(z)$ by truncating the product with a finite number of poles. The
resulting rational function can then be interpreted as the transfer
function of a finite-dimensional system.

Notice that the procedure used to obtain $\tilde T(z)$ does not
guarantee that $\tilde T(z) = T(z)$, although this is indeed the case
for the example above. In general the procedure above may not capture
some of the properties of the system. For example, if we added another
delay with no feedback in sequence, we would obtain an additional phase
factor dependent on $z$, while still satisfying the desired properties.
For a SISO system satisfying the unitarity constraint with the same
root-pole pairs, a phase dependence $ T(z) = \tilde T(z) e^{-\alpha z}$
($\alpha\ge0$) is actually the most general modification we might
need. For a system satisfying the same conditions but having $N$ input
and output ports, the $ e^{-\alpha z}$ term will be replaced by a
similar singular function which we will refer to throughout as the \emph
{singular} term (see Section~\ref{section:factorization_theorem} for a
discussion). An illustration of $T(z)$ and the transfer function
resulting when an additional delay is augmented to the system is
illustrated in Figure~\ref{fig:eg_complex}.

Notice how the augmented delay substantially changes some of the
properties of the transfer function in the complex plane. For instance,
when $\Re(z) \to-\infty$, the transfer function in Figure~\ref{fig:eg_complex}(a) approaches a constant, while that in Figure~\ref{fig:eg_complex}(b) diverges. In Section~\ref{section:singular} we will
be able to utilize the difference in behavior of transfer functions to
determine whether they incorporate a nontrivial everywhere analytic
term, like the exponential resulting from a time delay.

We can check to see if such a factor is needed in the factorization above.
Assuming that $\lim_{\Re(z) \to-\infty} \tilde T(z) = C \ne0$ (see
Appendix \ref{section:limit_argument}), we can take $\lim_{\Re(z) \to
-\infty} T(z)$ to check if the additional phase factor is present in
$T(z)$. We see $\lim_{\Re(z) \to-\infty} T(z) = -1/r$, which shows in
our example that no such additional term is needed, and therefore
$\tilde T(z) = T(z)$.
It can be confirmed numerically that the values of $T(z)$ and $\tilde
T(z)$ agree.

In the rest of the paper, we will show how a similar procedure can be
applied more generally to multiple-input and multiple-output (MIMO) systems.


\section{Factorization theorem and implications for passive linear systems}

\label{sec:fact_thm_implications}

Certain kinds of matrix-valued functions can be factorized using the
Potapov factorization theorem. A more general factorization
theorem is discussed in Appendix~\ref{section:potapov}. Here we discuss
a special case useful for our application. In Section~\ref{section:factorization_theorem} below we show how this special form can
be obtained using the theorems in Appendix~\ref{section:potapov}.

\emph{theorem}.
Let $T(z)$ be a meromorphic matrix-valued function satisfying
$T(z) T^{\dagger}(z) = I$ for $z\in i\mathbb{R}$ and bounded on the right
half-plane $\mathbb{C}^+$. Then we can represent $T$ with the factorization
%
\begin{align}
\label{equ:passive_potapov_fact} T (z) = U B(z) S(z),
\end{align}
where
%
\begin{align}
\label{equ:B_and_S} B(z) &= \prod_{k=1}^\infty
B_k(z), \qquad S(z) = \overset{\curvearrowright} {
\int_0^\ell} e^{-z K(t)\,dt}.
\end{align}
In the product, each of the terms $B_k$ has the form
%
\begin{align}
\label{equ:B_k} B_k(z) = V_k %
\begin{pmatrix}
e^{i \phi_k} \frac{z-\lambda_k }{z + \overline{\lambda_k} } I_{p_k\rq{}
} & 0 \\
0 & I
\end{pmatrix} %
V_k^{-1}
=I-P_k + A_k(z)P_k.
\end{align}
The $B_k$ terms are the Blaschke-Potapov factors written in the
$s$-plane formalism (see Appendix~\ref{section:cayley}),
and
$U$ and the $V_k$'s are some unitary matrices.
We use the symbol $\overset{\curvearrowright} {\int} $ to refer to
the \emph{multiplicative integral}, or product integral. The integral
above is given by
%
\begin{align}
\label{equ:mult_int_singular} S(z)= \lim_{\max\Delta t_j \to0} e^{-z K(t_0)(t_1-t_0)}
e^{-z K(t_1)(t_2 - t_1)} \cdots e^{-z K(t_{n-1})(t_n-t_{n-1})},
\end{align}
where we take $0 = t_0 \le t_1\le\cdots\le t_n = \ell$.
In the integral, $K(t)$ is a summable non-negative family of Hermitian
matrices with $\text{tr} [K(t)]=1$ on some interval $[0,\ell]$ for some
$\ell$.
The $P_k$ is an orthogonal projection, $A_k(z)= \frac{z-\lambda_k }{z +
\overline{\lambda_k} } e^{i \phi_k}$, and
$e^{i \phi_k}= \frac{| 1-\lambda_k^2| }{1-\lambda_k^2} $ is a phase
factor.

Notice that each Blaschke-Potapov factor has a zero at $\lambda_k$
and pole at $-\overline{\lambda_k}$.
We will refer to the terms $B(z)$ and $S(z)$ above as the Potapov
product and the singular term, respectively. Both $B(z)$ and $S(z)$ are
inner functions. A function $F(z)$ is said to be \emph{inner} if it
satisfies the unitarity condition $F(z) F(z)^{\dagger}= I$ for $z \in i
\mathbb{R}$ and is contractive on the right half-plane (\textit{i.e.} $F(z)
F^{\dagger}(z) \le I$ for $z \in\mathbb{C}^+$).
When $B(z)$ is a finite product, the $V_k$ terms can be redefined so
that the phase factors $\phi_k$ can be absorbed into the unitary matrix $U$.

\subsection{Obtaining the special case of the factorization theorem for
passive systems}
\label{section:factorization_theorem}

In this section, we will show how the theorems in Appendix~\ref{section:potapov} are
used to obtain the special form we will use.

First, notice we assumed above that $T(z)$ is a meromorphic
matrix-valued function unitary for $z\in i\mathbb{R}$ and bounded on
$\mathbb{C}^+$. These assumptions hold for the system described in
Section \ref{section:problem_characterization}. By Appendix~\ref{section:max}, it follows that $T(z)$ is an inner function on $\mathbb
{C}^+$.

We are now in a position to apply the Potapov factorization theorem
stated in Appendix~\ref{section:potapov}. Setting $J=I$ in the theorem
and applying the Cayley transformation (discussed in Appendix~\ref{section:cayley}) to map the unit disc in the theorem to the right
half-plane, the theorem implies that $T(z)$ has the form
%
\begin{align}
\bar B_\infty(z) \bar B_0(z) S(z),
\end{align}
where $\bar B_i(z) = B_i ( \frac{z-1}{z+1}  )$ for $i=0,\infty
$ and $S(z)$ results from a similar transformation on the
multiplicative integral in Eq. (\ref{equ:omega_potapov}). We will only
need to write the expression for $\bar B_0(z)$ explicitly. After some
algebra due to changing the domain from $\mathbb{D}$ to $\mathbb{C}^+$,
one obtains the form in Eq. (\ref{equ:B_k}).

Because $T(z)$ has no poles on $\mathbb{C}^+$, the term $\bar B_\infty
(z)$ is trivial.
Notice that $S(z)$ resulting in our discussion may only have poles only
for $z \in i \mathbb{R}$, because of the form of the integrand of the
multiplicative integral of the Potapov factorization.

$T(z)$ has no poles for $z \in i \mathbb{R}$. Therefore $S(z)$
is analytic everywhere.
The term $S(z)$ is unitary for $z \in i\mathbb{R}$ since $\bar B_0(z)$
and $T(z)$ are also unitary for $z \in i \mathbb{R}$.
We also note that $S(z)$ is contractive on $\mathbb{C}^+$, and refer
the reader to Potapov's paper \cite{Potapov1955} for details (in
particular, notice detaching single Blaschke-Potapov term from a
contractive function with the same zero results in a contractive function).
Since $S(z)$ is an entire contractive function unitary on $i \mathbb
{R}$, we can apply the second theorem from Appendix~\ref{section:potapov},
giving the expression in Eq.~(\ref{equ:singular_factorization}),
%
\begin{align}
S(z) = S(0) \overset{\curvearrowright} {
\int_0^\ell} e^{-z K(t)\,dt}.
\end{align}
If we wish, we can re-define the $V_k$ terms in $\bar B_0(z)$ so that
$\bar B_0(z) S(0) S(z) = U \tilde B_0 (z) S(z)$ (the convergence
criterion depends only on the zeros of the product).
Finally, we drop the tilde to obtain the form $T(z) = U B(z) S(z)$ in
Eq. (\ref{equ:passive_potapov_fact}).



\subsection{Interpretation as cascaded passive linear network}

We remark that the Blaschke-Potapov factorization of an inner
function can be interpreted as a limiting case of a system of
beamsplitters, feedforward delays, and cavity modes.

\begin{figure}
    \centering
    \includegraphics[width=0.2\textwidth]{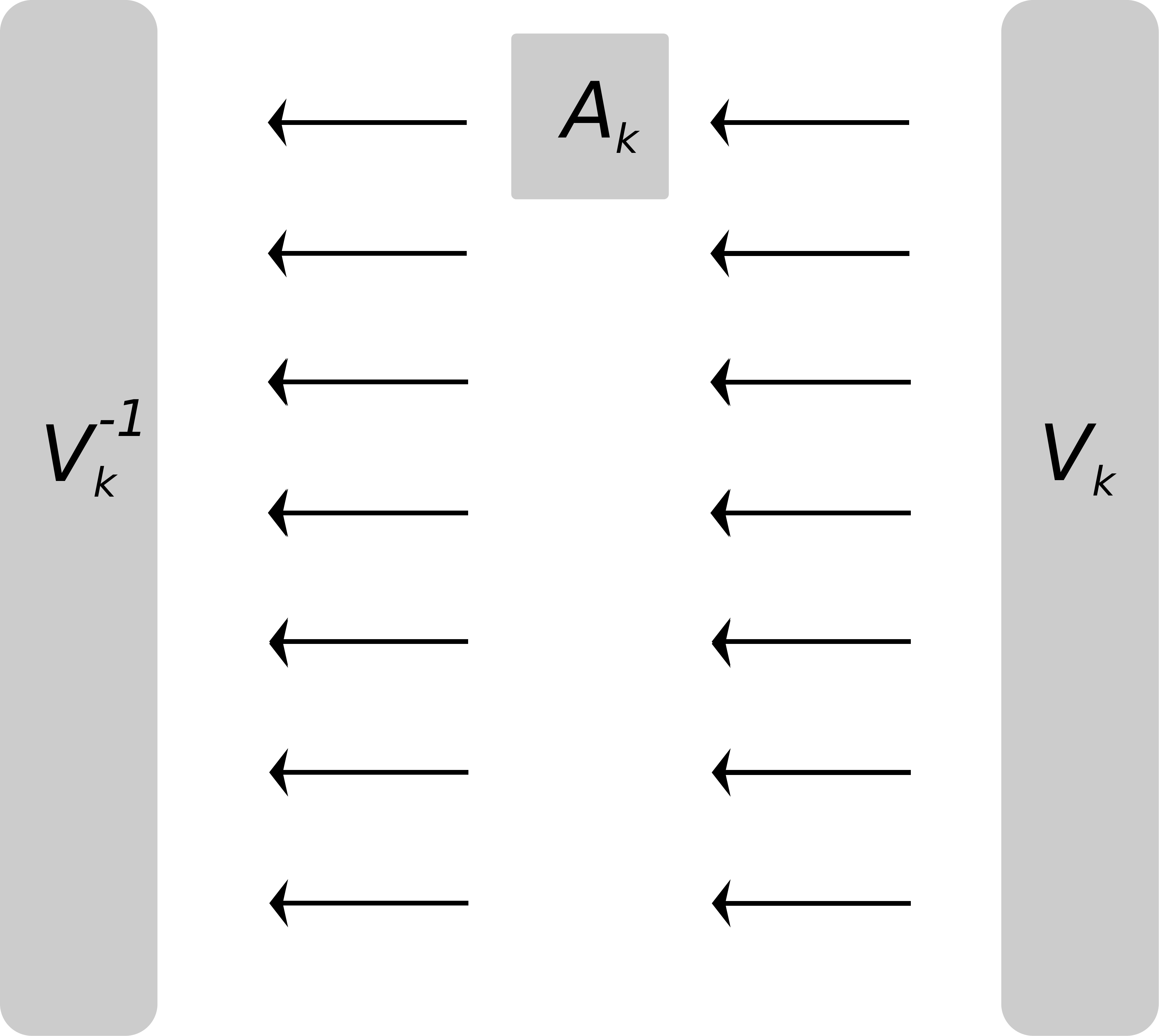}
\caption{\textbf{A physical interpretation of a single Blaschke-Potapov term.} Here the $V_k$
represents a beamsplitter and the $A_k$ represents a passive SISO system with a single degree of freedom.
The zero-pole interpolation generates a series of factors of this form in a cascade.}\label{fig:MIMO_Component}
\end{figure}

First, we will interpret the Blaschke-Potapov product in the optical setting.
Each unitary matrix appearing in the factorization can be interpreted
as a generalized beamsplitter.
Each Blaschke factor with zero $\lambda_k$ has the form $B_k(z)$ of
Eq. (\ref{equ:B_k}) in Section~\ref{section:factorization_theorem}. We
can interpret $A_k(z)$ in Eq. (\ref{equ:B_k}) as the transfer function
of a single cavity mode.
The location of the zero $\lambda_k$ in the complex plane determines
the detuning and linewidth of the mode.
The modes resulting from the Blaschke-Potapov product can be visualized
as a sequence of components of the form portrayed in Figure~\ref{fig:MIMO_Component}.

\begin{figure}
    \centering
    \includegraphics[width=0.2\textwidth]{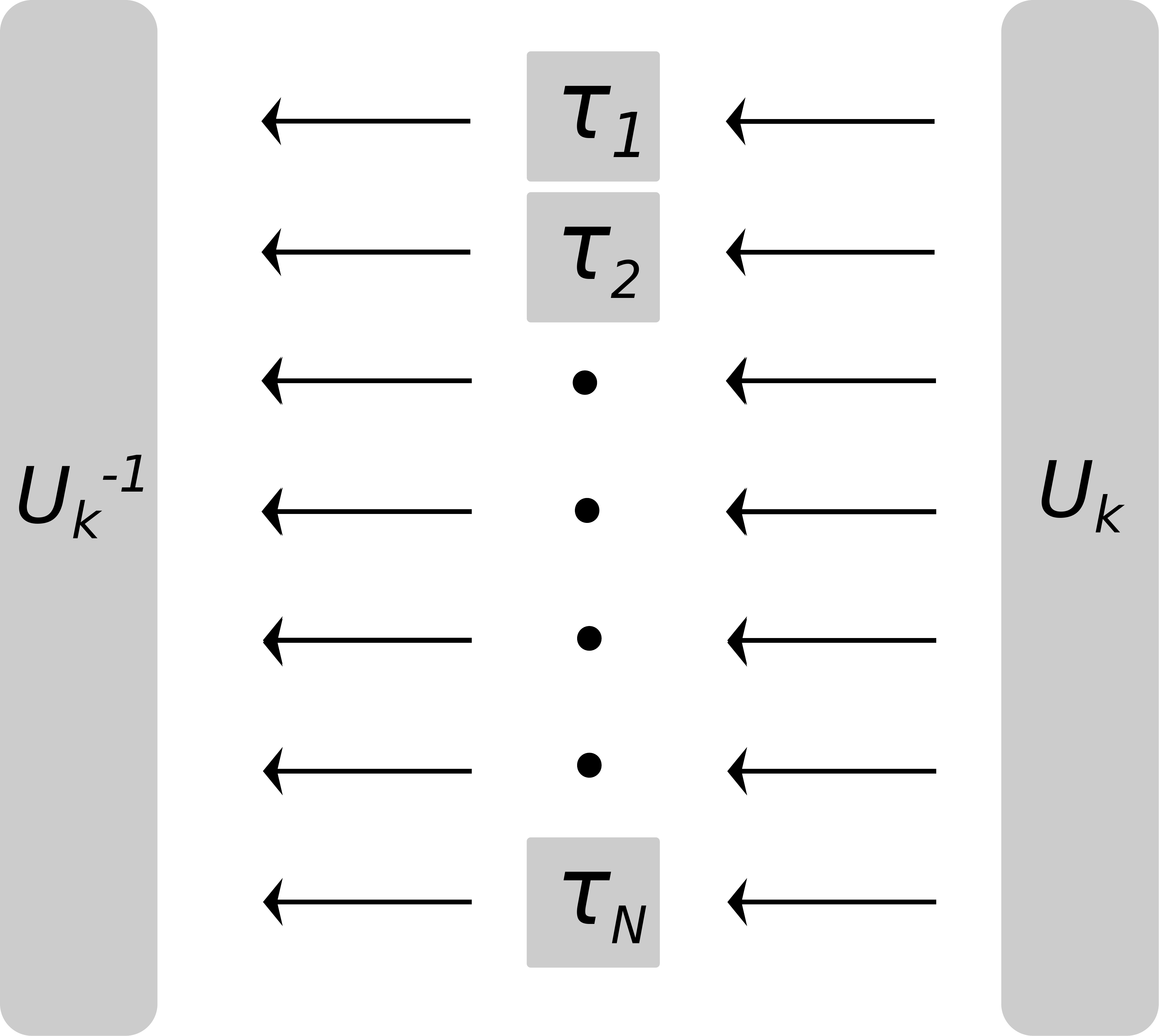}
\caption{\textbf{A physical interpretation of  a component resulting in an approximation of the multiplicative integral.}
Here the $U_k$ represents a beamsplitter and each $\tau_j$ ($j = 1,\ldots,N$) represents a
delay of that duration.}\label{fig:Time_Component}
\end{figure}

Next, we shall interpret the singular term represented as the
multiplicative integral in Eq.~(\ref{equ:mult_int_singular}) of
Section~\ref{sec:fact_thm_implications}. Approximating the
multiplicative integral over intervals $\Delta t_k = t_{k+1}-t_k$, we
obtain a product of terms that can each be represented by
%
\begin{align}
\label{equ:delay_term} e^{-z K(t_k)(t_{k+1} - t_k)} = e^{-z K(t_k)\Delta t_k} = U_k
e^{-zD_k} U_k^{\dagger}.
\end{align}
Here $K(t_k) \Delta t_k \ge0$ and $U_k$ are some unitary matrices
resulting in the diagonalization $K(t_k) \Delta t_k= U_k D_k U_k^{\dagger}$.
Each term of the form (\ref{equ:delay_term}) can be interpreted as a
component consisting of parallel feedforward delays inserted between
two unitary components as illustrated in Figure~\ref{fig:Time_Component}.
The sum of the delays across the parallel ports for each such term is
given by
$\text{tr} ( K(t_k) \Delta(t_k)) = \Delta t_k$.
It is possible to further approximate the feedforward delays in the
factorization with modes, but we will refrain from doing this here for
conceptual clarity


\section{Approximation Procedure -- Zero-Pole Interpolation}
\label{sec:zero-pole-interpolation}

In order to reconstruct an approximation for the transfer function
$T(z)$ using only a finite number of modes, we will use a two-step procedure.
The first step consists of finding the zero-pole pairs in a region of
interest. The second step consists of examining the numerical values of
the transfer function near the zeros or poles to obtain the correct
form of each of the Blaschke-Potapov terms, which are determined up to
a constant unitary factor.
The product of the resulting terms will equal a truncated version of
the Blaschke-Potapov product discussed in Section~\ref{section:factorization_theorem}, and will approximate the transfer
function in the region of interest.

We will take a transfer function $T$ and obtain an $M$-dimensional
approximation by identifying appropriate factors for a Blaschke-Potapov product.
It is possible that the transfer function may have a nontrivial
singular component (\textit{i.e.} a nontrivial everywhere analytic term) as
discussed in Section~\ref{section:factorization_theorem}, in which case
the zero-pole interpolation may not reproduce a converging sequence of
approximations to the given transfer function $T(z)$. In this section
we assume that the singular term is trivial or otherwise unimportant.
In any case we determine $U$ in Eq. (\ref{equ:passive_potapov_fact}) of Section~\ref{section:factorization_theorem} using $T(0)$.

A trivial example when the above approach might fail is a delay with no
feedback at all. In this case, there are no poles to evaluate and the
method fails. When a transfer function is entirely singular, or when
its singular component cannot be neglected, a different approach will
be needed, such as using the Pad\'{e} approximation. This is discussed
in Appendix~\ref{section:pade_approximation}.

\subsection{Identifying mode location}
\label{section:contour}

We remind the reader that we take our coordinate system in the $s$-domain.
We assume for simplicity that we are interested in the behavior of the
system near the origin. However, our procedure can be used to obtain
approximations of the given transfer function for arbitrary regions in
the $s$-plane.
In order to identify the appropriate modes, we find roots of the
transfer function of the full system, $\lambda_1,\ldots,\lambda_M$
(with corresponding poles $p_1,\ldots,p_M$). Each root will represent a
\lq\lq{}trapped\rq\rq{} resonant mode. In general, there will be
infinitely many such roots in the full system, so it is important to
have a criterion for selecting a finite number of roots. Each root will
have an imaginary part, which will correspond to the frequency of the
mode, and a real part, which is linked to the linewidth of the mode.
One criterion might be to select root whose imaginary part falls in
some range $[-\omega_{\max},\omega_{\max}]$, so that the approximation is
valid for a particular bandwidth.
This approximating system may be improved by increasing the maximum
frequency, $\omega_{\max}$.
As the number of zero-pole pairs increases, the quality of the
approximation increases, but in addition the approximated system will
incur a greater number of degrees of freedom.



Luckily there is a well-known technique that can be used based on
contour integration developed in \cite{Delves1967}.
This algorithm runs in a reasonable time and can essentially
guarantee that it does indeed find all of the desired points.
The latter point is an important feature that most typical
root-finding algorithms do not have because they do not utilize the
properties of analytic functions.
For details about a more polished algorithm see \cite{Kravanja2000}.
Methods of this kind require a contour in the complex plane as the
input in which the roots of the function will be found. This contour
may be, for example, a rectangle in the complex plane. In practice we
may make use of symmetries in the system and the known regions where
poles and zeros are located.

\begin{figure}
    \centering
    \includegraphics[width=0.4\textwidth]{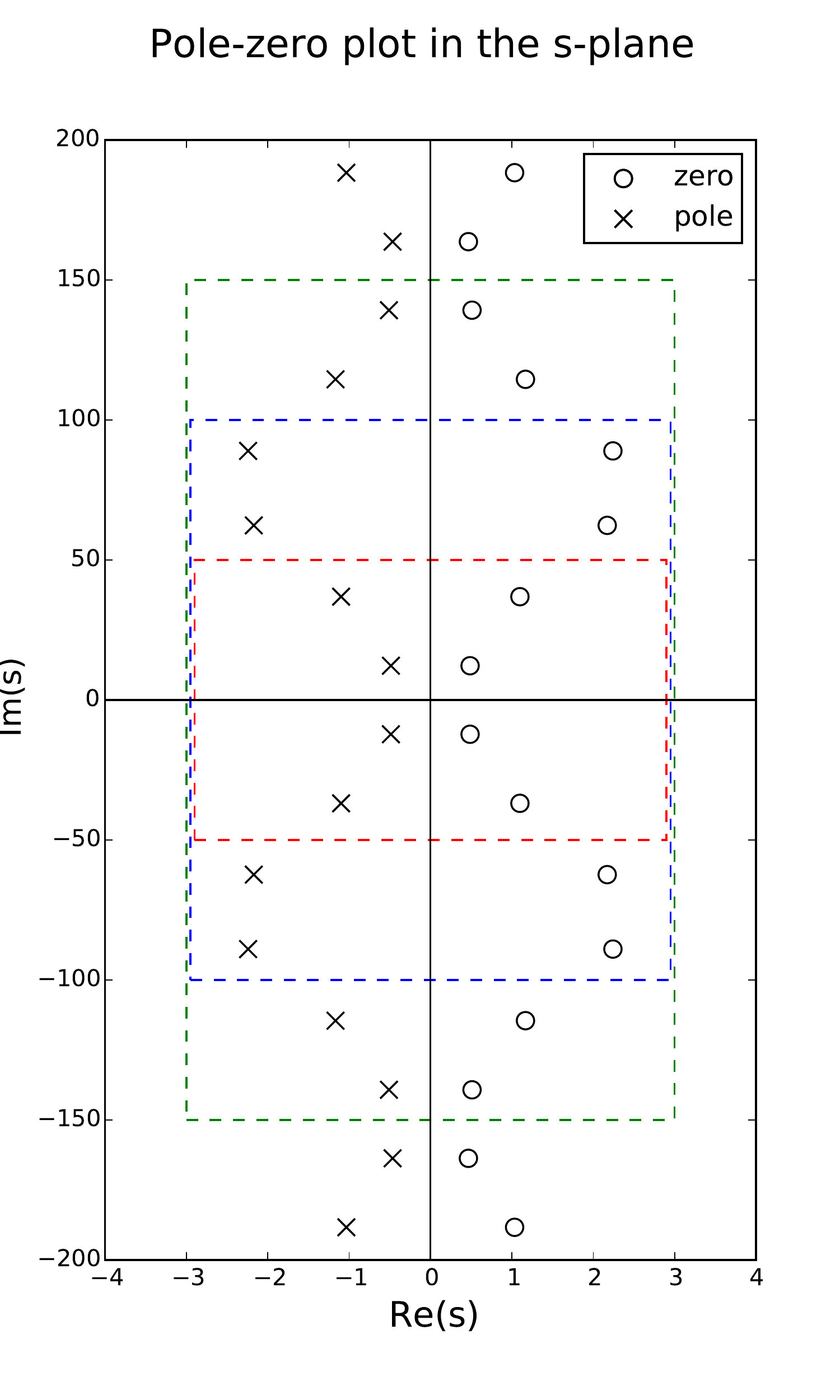}
\caption{\textbf{Sketch of root-finding procedure.} This plot illustrates the locations where root-pole pairs
may be found. As the contours grow and includes more root-pole pairs, the quality
of the approximation improves. The plot in this figure is
based on Example 2 from Section~\ref{sec:zero_pole_fail}.}\label{fig:scatter_plot}
\end{figure}

In Figure~\ref{fig:scatter_plot} we illustrate the step of our
procedure for finding roots or poles. The various contours in dashed
lines represent areas where roots and poles will be found. Notice in
this plot that the roots and poles lie along a strip close to the
imaginary axis.
This is a typical feature of highly resonant systems (\textit{i.e.}
effective cavity modes have a long lifetime) since the real part of
each pole in the system corresponds to the exponent of decay of each
mode. The system illustrated Figure~\ref{fig:scatter_plot} originates
from Example 2 in Section~\ref{sec:zero_pole_fail}.
If the maximum possible real part of each root is determined for the
system of interest, a computational advantage can be gained since the
contour does not need to be extended beyond that value.

\subsection{Finding the Potapov projectors}

The procedure we use assumes that the given transfer function $T(z)$
has a specific form guaranteed by the factorization theorem (see
Eq. (\ref{equ:passive_potapov_fact}) in Section~\ref{section:factorization_theorem}). For the purposes of this section, we
neglect the contribution due to the singular term (the $S(z)$ in
Eq. (\ref{equ:passive_potapov_fact})).
This procedure is similar to the zero-pole interpolation discussed
in \cite{Ball1990}. We handle the singular term separately, as we will
discuss in Section~\ref{section:singular}.

We introduce an inductive procedure for this purpose. Each step
will involve extracting a single factor of the Blaschke-Potapov
product. We suppose the full transfer function being approximated is
$T(z)$, and that it has a pole at $p$. Based on the form of the
Blaschke-Potapov factors, we can separate the transfer function into
the product
%
\begin{align}
\label{equ:inductive_1} T(z) = \tilde T(z) \biggl( I-P+ P \biggl( \frac{z+\overline p}{z-p}
\biggr) \biggr).
\end{align}
The $P$ is in general the orthogonal projection matrix onto the
subspace where the multiplication by the Blaschke factor takes place.
We wish to extract the $P$ given the known location of the pole $p$,
which we assume to be a first-order pole for simplicity.
We also assume for simplicity that $P$ is a rank one projection, and so
it can be written as the product of a normalized vector
\[
P = v v^{\dagger}.
\]
The simplifying assumptions above have been sufficient for the systems
we inspected, and could be easily removed.
Rewriting, we obtain the relationship
%
\begin{align}
T(z) (z-p) = (z-p) \tilde T(z) (I-P) + ( {z+\overline p} ) \tilde T(z) v
v^{\dagger}.
\end{align}
Now take $z\to p$.
We assumed that $T(z)$ has a first-order pole at $p$, so $\tilde T(z)$
will be analytic at $p$.
Therefore, the first term on the right hand side goes to zero. Taking
$L \equiv\lim_{z \to p} T(z) \*(z-p) $, we get
%
\begin{align}
L = ( {p+\overline p} ) \tilde T(p) v v^{\dagger}.
\end{align}
Since we assumed that $P$ is a rank one projector, have obtained an
expression where $L$ must also be rank one.
In order to find $v$ we can simply find the normalized eigenvector
corresponding to the nonzero eigenvalue of $L$. This task may be done
numerically. Finally, we can find the $\tilde T(z)$ from Eq. (\ref
{equ:inductive_1}) above.

The procedure outlined above may be repeated for each of the $M$
desired roots of $T(z)$ to obtain a factorization
%
\begin{align}
\label{equ:inductive_k} T(z) = T_M(z) \prod_{k=1}^M
\biggl( I-P_k+ P_k \biggl( \frac{z+\overline
p_k}{z-p_k} \biggr)
\biggr).
\end{align}
We assume that the $T_M(z)$ is close to a constant in the region of
interest. This is exactly true in the case where the transfer function
$T$ has only the $M$ roots picked. We can approximate $T_M$ with a
unitary factor that can be determined from $T$ and the product in
Eq.~(\ref{equ:inductive_k}) evaluated at some point $z_0$ in the region
of interest.

The computer code for this procedure can be found on \cite{Tabak2016}.


\begin{figure}[b]
    \centering
    \includegraphics[width=0.6\textwidth]{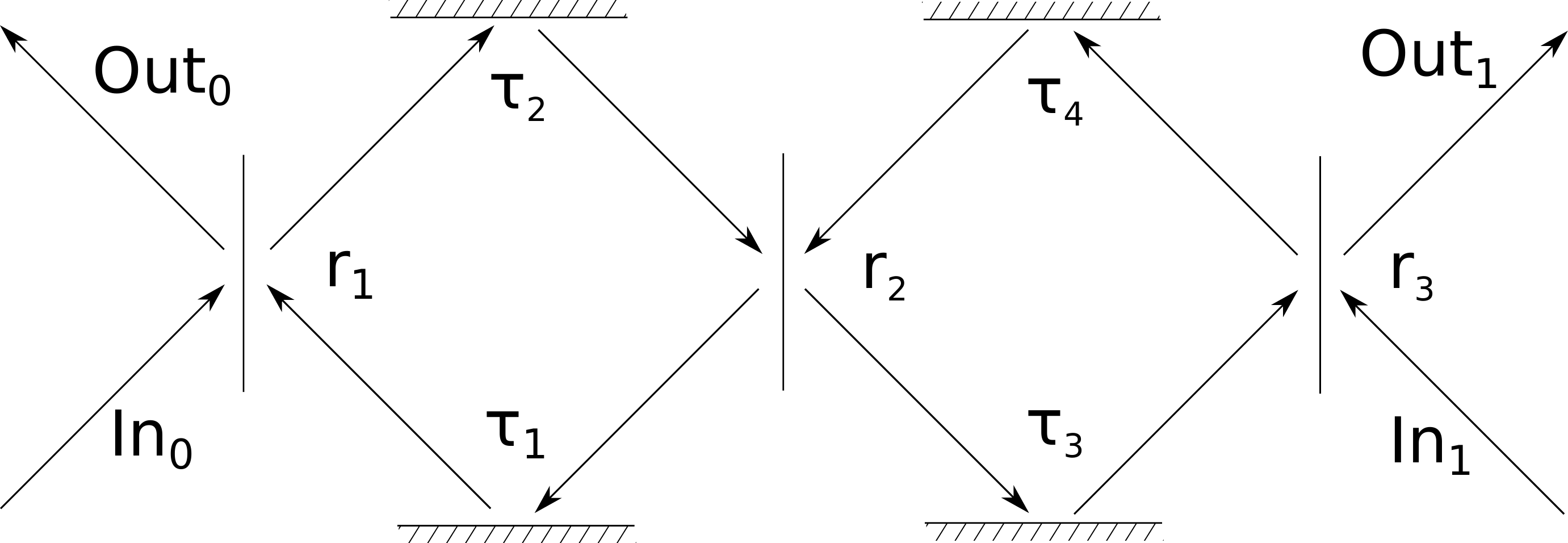}
\caption{\textbf{The network of Example 1 in Section~\ref{sec:zero_pole_success}.} For this network the zero-pole
interpolation produces accurate approximating functions because the singular term is trivial.}\label{fig:Example_5_pic}
\end{figure}

\begin{figure}
    \centering
    \includegraphics[width=1.0\textwidth]{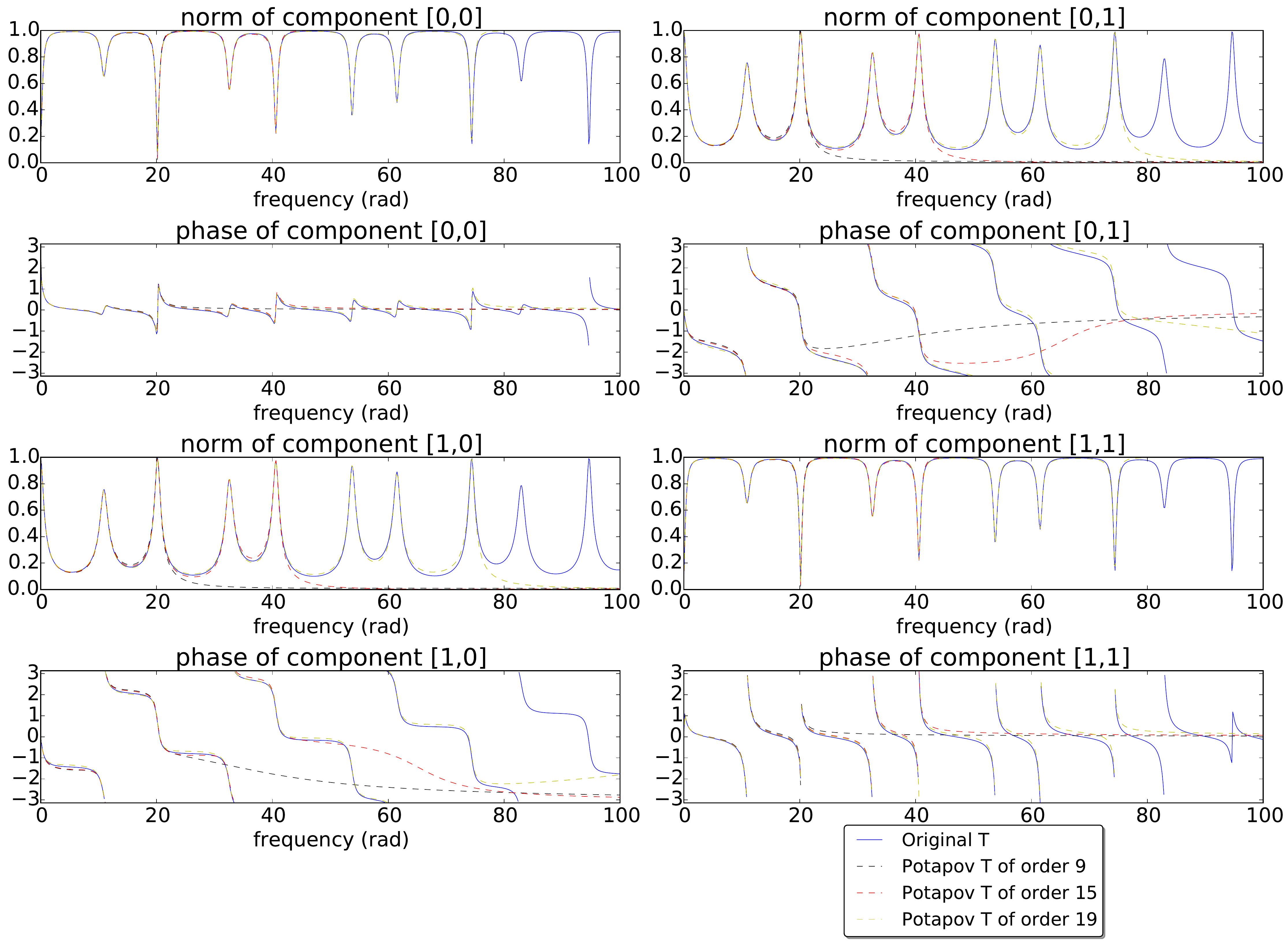}
\caption{\textbf{A plot of the transfer function $T$ of Example 1 from Section~\ref{sec:zero_pole_success}
and various approximated transfer functions generated by the zero-pole interpolation method.}
We illustrate the  components of $T(i \omega)$ for $\omega \ge 0$.
    We only include nonnegative $\omega$ because of the symmetry of both $T$ and its approximations.
     The zero-pole interpolation appears to converge to the correct transfer function as we add more terms.}\label{fig:Example_4_graph}
    \end{figure}

\section{Examples of zero-pole decomposition}
\label{section:examples}

In this section, we show two examples where we have applied the
zero-pole procedure.
The networks used for these examples are shown in Figures~\ref{fig:Example_5_pic} and \ref{fig:eg_4_2}. We plot the various
components of the transfer functions of these networks along $i \omega$
for $\omega\ge0$ in Figures~\ref{fig:Example_4_graph} and~\ref
{fig:Example_5_graph}, respectively.
Along both examples, we also plot several approximate transfer
functions determined by the zero-pole interpolation of Section~\ref{sec:zero-pole-interpolation}. The approximate transfer functions
correspond to a Blaschke-Potapov product that has been truncated to a
certain order. In both examples we see that as we increase the number
of terms, the approximation improves.
The first example illustrates the case when the zero-pole interpolation
converges to the correct transfer function. In the second example,
while the zero-pole interpolation appears to converge, the function to
which it converges deviates from the original transfer function. This
suggests that the singular term $S(z)$ in Section~\ref{section:factorization_theorem} makes a contribution for which the
zero-pole interpolation does not account. In Section~\ref{section:singular}, we discuss a condition for convergence and show how
the effects of the singular term may be separated from the rest of the
system. Figure~\ref{fig:Example_5_graph} also includes the transfer
function once the singular term has been removed, demonstrating
that the zero-pole approximations converge to that function.

\subsection{Example 1. Zero-pole interpolation converges to given
transfer function}
\label{sec:zero_pole_success}

The first example we discuss involves two inputs and two outputs.
Figure~\ref{fig:Example_5_pic} shows this network explicitly.
In Figure~\ref{fig:Example_4_graph} we see that the zero-pole
interpolation appears to converge to the correct transfer function. We
can check this by confirming that the $M_1$ is nonsingular, as we will
show in Section~\ref{section:singular}.

The matrices of Eq. (\ref{equ:delays_and_beamsplitter}) in Section~\ref{section:problem_characterization} are given by
%
\begin{align}
M_1 = %
\begin{bmatrix}
0&-r_1&0&0 \\
-r_2&0&t_2&0 \\
0&0&0&-r_3 \\
t_2&0&r_2&0
\end{bmatrix} %
, \qquad
M_2 = %
\begin{bmatrix}
t_1&0 \\
0&0 \\
0&t_3 \\
0&0
\end{bmatrix} %
,
\\
\\
M_3 = %
\begin{bmatrix}
0& t_1&0&0 \\
0&0&0&t_3
\end{bmatrix} %
, \qquad
M_4 = %
\begin{bmatrix}
r_1&0\\
0&r_3
\end{bmatrix} %
.
\end{align}

Here
$
\tau_1 = 0.1$,
$\tau_2 = 0.23$,
$\tau_3 = 0.1$,
$\tau_4 = 0.17$,
$r_1 = 0.9$,
$r_2 = 0.4$,
$r_3 = 0.8$.

\subsection{Example 2. Zero-pole interpolation fails to converge to
given transfer function}
\label{sec:zero_pole_fail}

\begin{figure}
    \centering
    \includegraphics[width=0.4\textwidth]{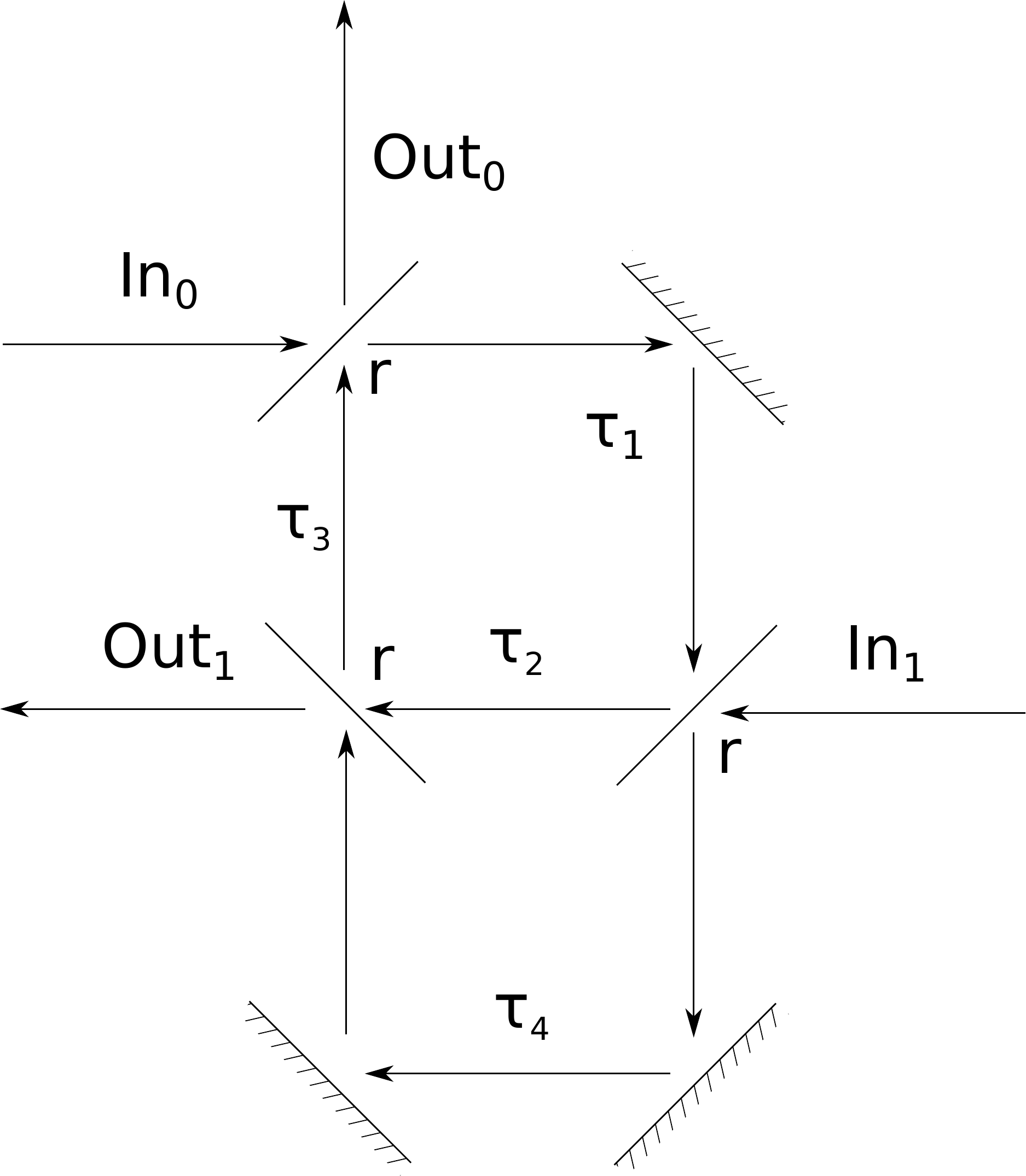}
\caption{\textbf{The network of Example 2 in Section~\ref{sec:zero_pole_fail}.} In this network the zero-pole
interpolation method would not be wholly applicable because the singular term is nontrivial.}\label{fig:eg_4_2}
\end{figure}

In the next example, we have two inputs and two outputs, as in the
first example of Section~\ref{sec:zero_pole_success}. However, the
design of the network is significantly different. The network for this
example is shown in Figure~\ref{fig:eg_4_2}. This example combines
elements of an interferometer and an optical cavity. In some regimes,
such as $\tau_4 \ll \tau_2$, the zero-pole decomposition yields a good
approximation for the transfer function. In general, however, the
singular component of the transfer function must be incorporated in
some other way.

The matrices of Eq. (\ref{equ:delays_and_beamsplitter}) in Section~\ref{section:problem_characterization} are given by
%
\begin{align}
\label{equ:M1_not_work} M_1 = %
\begin{bmatrix}
0 &   0 &   -r &   0 \\
r &   0 &   0  &   0 \\
0 &   r &   0  &   t \\
t &   0 &   0  &   0
\end{bmatrix} %
,
\qquad M_2 = %
\begin{bmatrix}
t&0 \\
0&t \\
0&0 \\
0&-r
\end{bmatrix} %
,
\\ \\
M_3 = %
\begin{bmatrix}
0&0&t&0 \\
0&t&0&-r
\end{bmatrix} %
, \qquad
M_4 = %
\begin{bmatrix}
r&0 \\
0&0
\end{bmatrix} %
.
\end{align}

Here
$
\tau_1 = 0.1$,
$\tau_2 = 0.039$,
$\tau_3 = 0.11$,
$\tau_4 = 0.08$,
$r = 0.9$.

\begin{figure}
    \centering
    \includegraphics[width=1.0\textwidth]{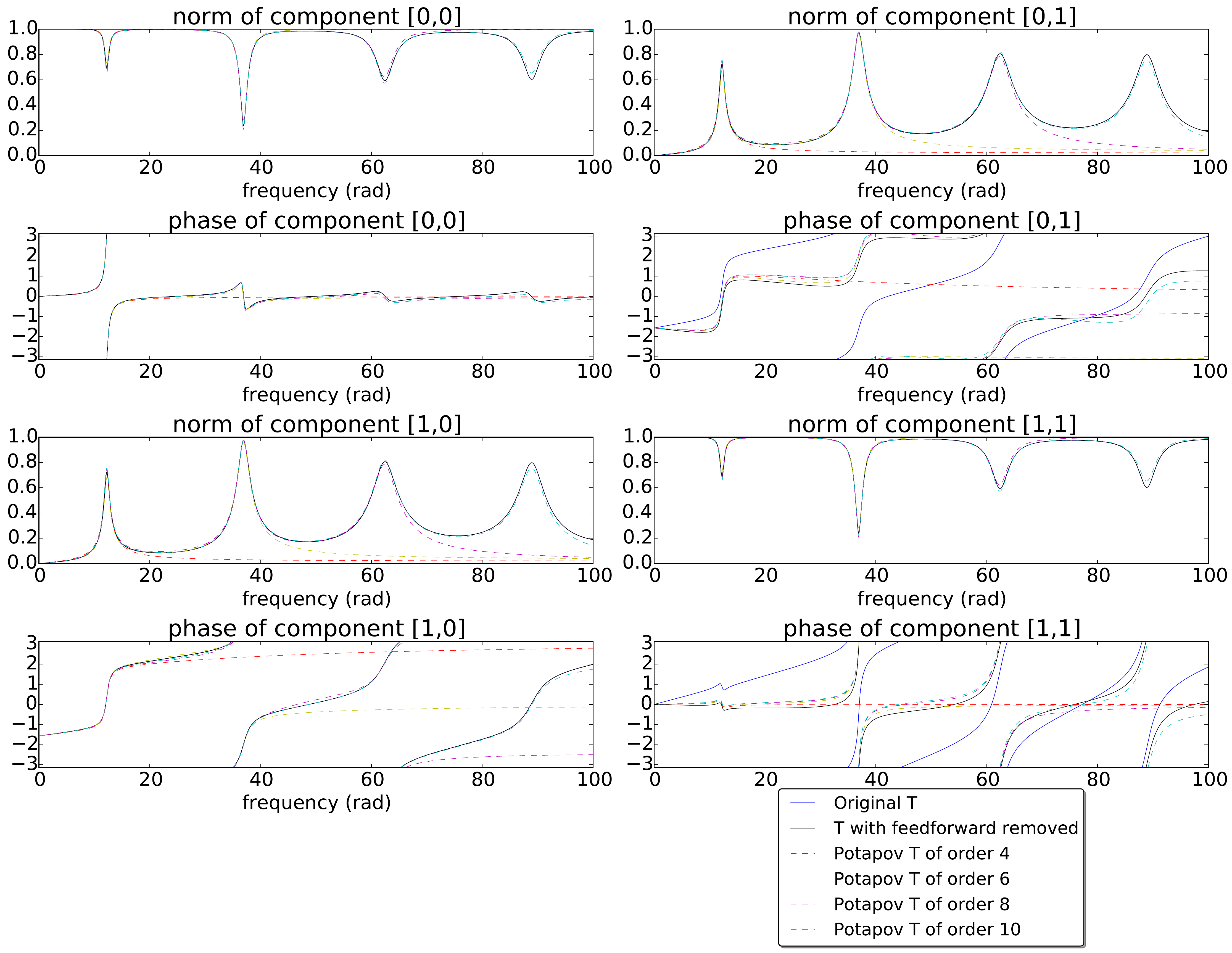}
\caption{\textbf{A plot of the transfer function $T$ of Example 2 from Section~\ref{sec:zero_pole_fail}
and various approximated transfer functions generated by the zero-pole interpolation method.} We also include the transfer
function resulting when the singular term is removed, to which the approximated functions converge.
We illustrate the  components of $T(i \omega)$ for $\omega \ge 0$.
    We only include nonnegative $\omega$ because of the symmetry of both $T$ and its approximations.
     See Section~\ref{sec:zero_pole_fail} for further details.}\label{fig:Example_5_graph}
\end{figure}

The important differentiating feature
from the previous example of Section~\ref{sec:zero_pole_success}
is that the singular term for the transfer function of this network is
nontrivial.
This can be seen when examining the resulting transfer functions from
the zero-pole interpolation, which are shown in Figure~\ref{fig:Example_5_graph}.
In Section~\ref{section:singular}
we will show that this condition can be checked by observing that the
$M_1$ in Eq. (\ref{equ:M1_not_work}) is singular.

In Figure~\ref{fig:Example_5_graph}, we see that the zero-pole
interpolated transfer functions deviate from the true transfer function
in the $(0,1)$ and $(1,1)$ phase components. This demonstrates how in
general it is important to consider the singular function. On the other
hand, for the systems in consideration it is possible to separate the
Blaschke-Potapov product from the singular term, which corresponds to
feedforward-only components, as discussed in Section~\ref{section:separation_1}. In black we graph the transfer function
components resulting once the feedforward-only components have been
removed. Up to a unitary factor, this function is equal to the infinite
Potapov product. We see that the approximated transfer functions from
the zero-pole interpolation converge to this function.

%

\section{The Singular Term}
\label{section:singular}

In this section, we examine the factorization of the transfer function
 given in Eq. (\ref{equ:passive_potapov_fact}) in Section~\ref{section:factorization_theorem}.
In the form of the fundamental theorem by Potapov that we obtained, we
had an infinite product of Blaschke-Potapov factors and a singular term.
Although the zero-pole decomposition allowed us to extract the
Blaschke-Potapov factors, it gave us no information regarding the
singular term. In some systems, it may be crucial to include the
singular term to obtain a good approximation of the system. To learn
about this term, we will need a different method.

In this section, we give a condition for the singular term to be
trivial. This condition can then be specialized to the network from
Section~\ref{section:problem_characterization}. Based on this
condition, we can develop a method to explicitly separate the network
 described by Eq. (\ref{equ:delays_and_beamsplitter}) in
Section~\ref{section:problem_characterization} into the Potapov product
and the singular term.

\subsection{Condition for the multiplicative integral term to be trivial}

We examine the form of the singular term in the factorization theorem
and notice that its determinant becomes large when $\Re(z) \to-\infty$.
To avoid mathematical details, we will assume here that the
Blaschke-Potapov product $\prod B_k(z)$ is well-behaved in the limit
$\Re(z) \to-\infty$ in the sense that the limit of the product
converges (to a nonzero constant). Justification for this assumption is
discussed further in Appendix~\ref{section:limit_argument}. We have the
following observations.

\emph{Observation.}
If $\lim_{\Re(z) \to-\infty} T(z)$ is a constant,
then the multiplicative integral in Eq. (\ref
{equ:passive_potapov_fact}) of Section~\ref{section:factorization_theorem} is a constant.

This follows from the properties of the multiplicative integral defined
in Eq. (\ref{equ:mult_int_def}) of Appendix~\ref{section:defs}.

\emph{Observation.}
In particular, for the transfer function $T(z)$ in Eq. (\ref
{equ:passive_TF}) of Section~\ref{section:problem_characterization},\linebreak[4]
$\lim_{\Re(z) \to-\infty} T(z)$ is a constant if and only if $M_1$ in
Eq. (\ref{equ:passive_TF}) is full-rank. This gives a sufficient
condition for when the zero-pole expansion converges exactly.

To obtain this result, it is enough to consider the term $(E(-z) -
M_1)^{-1}$ in the limit $\Re(z) \to-\infty$.

%

The above observations can be seen in the two examples discussed in
Section~\ref{section:examples}. In Example~1, the $M_1$ matrix is
full-rank, while in Example 2 it is not.

\subsection{Maximum contribution of singular term}

For many applications we anticipate that we may be able to drop the
contribution of the singular term altogether. One example is an optical
cavity in certain regimes. If the lifetime of the modes in the cavity
is long in comparison to the delays in the system, we would expect the
delays to be less significant. We would like to be able to provide a
justification for when it is acceptable to neglect the singular term.

First, we will obtain the maximum value for $\ell$ necessary in the
multiplicative integral appearing in Eq. (\ref{equ:B_and_S}) of
Section~\ref{section:factorization_theorem}.
This is an important result because it tells us that the lengths of the
delays themselves determines the greatest contribution of the singular function.

\emph{Remark.}
 To apply the factorization in Eq. (\ref
{equ:passive_potapov_fact}) of Section~\ref{section:factorization_theorem} to the transfer function in Eq.~(\ref
{equ:passive_TF}) of Section~\ref{section:problem_characterization}, it
suffices to take $\ell\le\sum_k T_k$.

This can be seen by noting the scaling of $\det[(E(-z) - M_1)^{-1}]$
in the limit $\Re(z) \to- \infty$.
%
%

The above bound occurs in the case of several delays feeding forward in
sequence.


We can give one condition under which the singular term can be dropped:
$| z |  \ll 1 / \ell$. Furthermore, crude estimates for the error can now
be found using the Taylor expansion of the exponential.

Intuitively, Potapov factors correspond to resonant modes while the
singular function corresponds to feedforward-only components.
With this interpretation, we see that the zero-pole interpolation
yields a transfer function close to the true transfer function when the
feedforward-only term can be neglected. We can interpret $\ell$ as an
upper bound on the duration of time the signal can spend being
fed-forward only. When $1/\ell$ becomes large with respect to the size
of the region of interest in the frequency domain, the feedforward-only
terms become unimportant.

\subsection{Separation of the Potapov product and the singular term in
an example}\label{section:separation_1}

In this section, we discuss how for a network of beamsplitters and
delays the Blaschke-Potapov product and the singular term of
Section~\ref{section:factorization_theorem} can be separated
explicitly. We will give a systematic procedure at least with the
simplifying assumption that the delays are commensurate (are rational
multiples of one another). In practice, one can always approximate
the delays to arbitrary precision with commensurate delays, resulting
in a large but sparse network.

\begin{figure}
    \centering
    \includegraphics[width=1.0\textwidth]{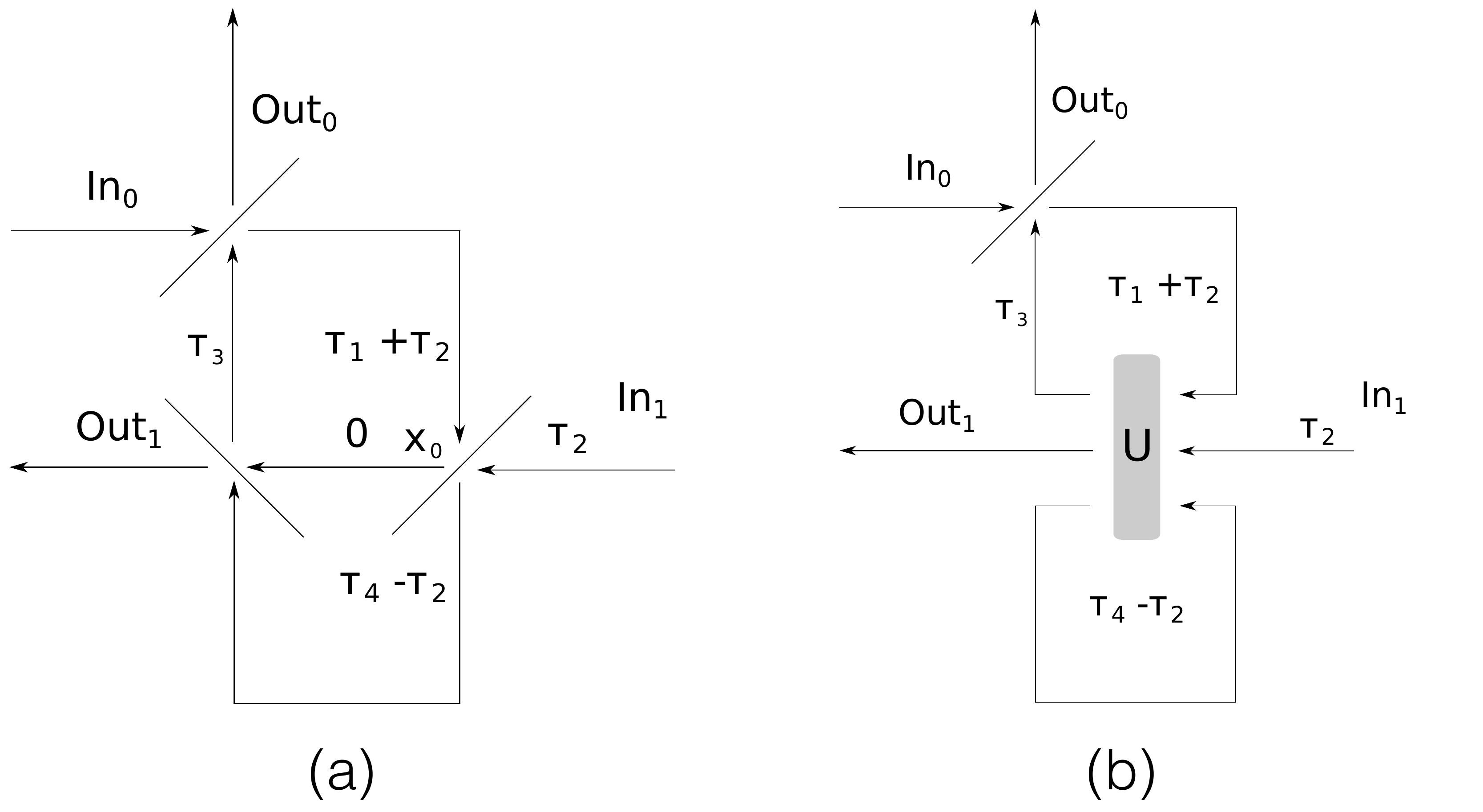}
\caption{\textbf{An illustration of one way the Potapov product and the singular function can be separated for
Example 2 in Section~\ref{sec:zero_pole_fail}.}  We notice that some of the parallel delays can be
commuted with a beamsplitter, forming the network in \textbf{(a)}.
The node labeled $x_0$  can then be eliminated,
forming the network in \textbf{(b)}
which now has a 3-input 3-output unitary component denoted by $U$.}\label{fig:separate_terms}
\end{figure}

To intuitively motivate the procedure we will use, we first
give a demonstration for the case of the example in Section~\ref{sec:zero_pole_fail}. In this network, extracting a single feedforward
delay is sufficient for obtaining the separation of the two terms we
desire. We will assume that $\tau_2 < \tau_4$. The important
observation is that a collection of $k$ parallel delays can be commuted
with a given a unitary component $U$ of $k$ ports. This is illustrated
in Figure~\ref{fig:separate_terms}(a).

Next, it becomes apparent in the new network shown in Figure~\ref{fig:separate_terms}(a) that one of the internal system nodes is
unnecessary, since it is followed by a delay of duration zero (call it
$x_0$). For this reason, $x_0$ can be eliminated from the network. In
the process, we can combine two of the unitary components preceding and
following $x_0$ to form a separate unitary component, illustrated in
Figure~\ref{fig:separate_terms}(b). The network depicted in Figure~\ref{fig:separate_terms}(b) can be decomposed into a feedforward-only
component followed by a network for which the $M_1$ matrix is invertible.
The feedforward-only component consists of the identity applied to
$\In_{0}$ combined in parallel with the addition of the
delay $\tau_2$ to $\In_{1}$ - that is, the feedforward
component only delays the input from port $\In_{1}$ by $\tau_2$.
The network following the feedback-only component
results from the exclusion of the delay $\tau_2$ in Figure~\ref{fig:separate_terms}(b).
Since its $M_1$ matrix is invertible, this network has trivial singular part.

\subsection{Systematic separation of Potapov product and analytic term
for passive delay networks}

Suppose we have a system given in the form of Eq. (\ref
{equ:delays_and_beamsplitter}) in Section~\ref{section:problem_characterization}. Our observation that in order for
the multiplicative integral term to be trivial we need $M_{1}$ to be
invertible suggests
that there may be a way to isolate the Potapov product term from the
remaining analytic function. We now present a systematic way of doing
this. For simplicity we will assume that all the delays are
comensurate. Without a loss of generality, we can write the system in
such a way that all the delays have equal duration, and therefore
$E(z)$ is a multiple of the identity.

The essential idea is to make a change of basis that allows the
elimination of a node at the expense of modifying the inputs in such a
way that a part of the analytic term can be extracted from the network.
We are interested in the case with $M_1$ not invertible. When this is
the case, we can find some change of basis represented by the matrix
$S$ such that
%
\begin{align}
M_1 = S J S^{-1},
\end{align}
where the $J$ is the Jordan decomposition of $M_1$ such that the
zero eigenvalue block is at the right bottom block of $J$.
We introduce $\bar x = S^{-1} x$ and rewrite the equation for $x$ in
Eq. (\ref{equ:passive_TF}) of Section~\ref{section:problem_characterization} as
%
\begin{align}
\bar x = J S ^{-1} E(z) S \bar x+ S ^{-1}M_2
x_{\mathrm{in}}.
\end{align}
Now, $\bar x_1 = [S^{-1} M_2 x_{\mathrm{in}}]_1$, which depends only on
the inputs.
The subscript here refers to the first component.
We can separate the dependence of the remaining coordinates of $\bar
x$. Denoting the last column of $S$ by $S_1$, the matrix of columns
excluding the last as $S_{\setminus1}$, and the matrix of rows of $J$
excluding the last as $J_{\slash1}$, we can write
%
\begin{align}
\bar x_{\setminus1} = J_{\slash1} S ^{-1} E(z) [S_1
\bar x_1 + S_{\setminus1} \bar x_{\setminus1} ]+ S
^{-1}M_2 x_{\mathrm{in}}.
\end{align}
We can now separate the network into two networks. The first network
takes the original inputs $x_{\mathrm{in}}$ and yields the outputs
%
\begin{align}
\tilde x_{\mathrm{in}} = J_{\slash1} S ^{-1} E(z)
S_1 \bigl[S^{-1} M_2 x_{\mathrm{in}}\bigr] +
S^{-1} M_2 x_{\mathrm{in}}.
\end{align}
Notice the first network is a feedforward network (\textit{i.e.} no signal feeds
back to a node from which it originated).
The second network takes the $\tilde x_{\mathrm{in}}$ as inputs and
yields the outputs
%
\begin{align}
\label{equ:H_setminus_1} \bar x _{\setminus1} = J_{\slash1} S ^{-1} E(z)
S_{\setminus1} \bar x_{\setminus1} + \tilde x_{\mathrm{in}}.
\end{align}
%
Using the simplifying assumption that the $E(z) = I e(z)$ is a multiple
of the identity, we obtain
%
\begin{align}
\label{equ:H_setminus_2} \bar x_{\setminus1} = \tilde J e(z) \bar x_{\setminus1} +
\tilde x_{\mathrm{in}},
\end{align}
where the $\tilde J$ is matrix resulting from dropping the last row
and column in $J$.
We see that Eq.~(\ref{equ:H_setminus_2}) has the same form to the
original equation for  $x$  in Eq. (\ref{equ:passive_TF}) in
Section~\ref{section:problem_characterization}. The difference is that
now the $\tilde J$ replaces $M_1$, and has one fewer zero eigenvalue.
Conveniently, $\tilde J$ is also in its Jordan normal form, so the
procedure can be repeated until the matrix ultimately replacing $M_1$
(call it $\tilde M_1$) has no zero eigenvalues left. In this case
$\tilde M_1$ is an invertible matrix, which is exactly the condition we
needed for the transfer function of the network to consist of only the
Potapov product and not the multiplicative integral.

A very simple example illustrating the intuition of our procedure is a
network where the internal nodes all feed forward in sequence.
Explicitly, take
%
\begin{align}
M_1 = %
\begin{pmatrix}
0 & 1 & 0 & 0 \\
0 & 0 & 1 & 0 \\
0 & 0 & 0 & 1 \\
0 & 0 &0 & 0
\end{pmatrix} %
.
\end{align}
Notice that this matrix is already in its Jordan-canonical form, so the
analysis becomes transparent. Also notice that all of the eigenvalues
of $M_1$ are zero, which implies that our procedure will extract all
the delays and collect them in the singular term.


\section{Relationship to the ABCD and SLH formalisms}
\label{section:ABCD_and_SLH}

In this section we demonstrate how the approximating system our
procedure designs is physically realizable. In particular we show how
to extract the \textit{ABCD} and SLH forms for a single term resulting in the
truncated Blaschke-Potapov product designed to approximate the transfer
function of the system.
Since the transfer function is equal to a product of such terms, we can
interpret the approximating system as a sequential cascade of
single-term elements of this form.

%
%
%
%
%
The state-space representation of the Potapov factor will have the form
%
\begin{align}
B(z) = C(Iz - A)^{-1} B+D.
\end{align}

To obtain the \textit{ABCD} model for a single Potapov factor, begin with the
following factor
%
\begin{align}
\label{equ:potapov_factor} B(z) = I-v v^{\dagger}+v v^{\dagger} \biggl(
\frac{z+\overline\lambda}{z-\lambda
} \biggr) = I + v v^{\dagger}\frac{\lambda+ \overline \lambda}{z-\lambda}.
\end{align}
%
In this instance we have also assumed that the orthogonal projector $P
= v v^{\dagger}$ has rank one.

There is some freedom in how the $B$ and $C$ matrices may be chosen. In
particular, one choice is also consistent with the form used for
passive components in the SLH formalism.
The \textit{ABCD} formalism is related to the SLH formalism in the following way
for a passive linear system.
%
\begin{align}
A = -\frac{1}{2}C^{\dagger} C - i \Omega, \qquad B =
-C^{\dagger} S, \qquad D=S.
\end{align}
In order to satisfy the above equations, we choose
%
\begin{align}
B = -\sqrt{-(\lambda+ \overline\lambda)}v^{\dagger}, \qquad C = \sqrt{-(
\lambda+ \overline \lambda)}v.
\end{align}
Finally, we can solve for the $\Omega$.
%
\begin{align}
\Omega=\biggl[ i\biggl(A - \frac{1}{2}(\lambda+ \overline\lambda) \biggr)
\biggr] =-\bigl[\Im(A)\bigr].
\end{align}
For the last equality, we use that $\frac{1}{2}(\lambda+ \overline
\lambda)$ is exactly the real part of the eigenvalue, and so cancels
exactly with the real part of $A$. The only remaining component is the
imaginary part of $A$, which is multiplied by $i$. Notice the $\Omega$
satisfies the condition of being Hermitian.

\section{Simulations in Time Domain}
\label{section:time_domain}

We translate our model into the \textit{ABCD} state-space formalism, as
discussed in Section~\ref{section:ABCD_and_SLH}. Doing this allows us
to run a simulation in the time domain.
Notice that for linear systems this approach suffices for finding the
dynamics in the time domain.
We can apply an input field at some frequency and record the output.
The relationship between the inputs and outputs at the steady-state
will correspond to the value of the transfer function at the
appropriate frequency.

\begin{figure}
	\centering
	\includegraphics[width=0.9\textwidth]{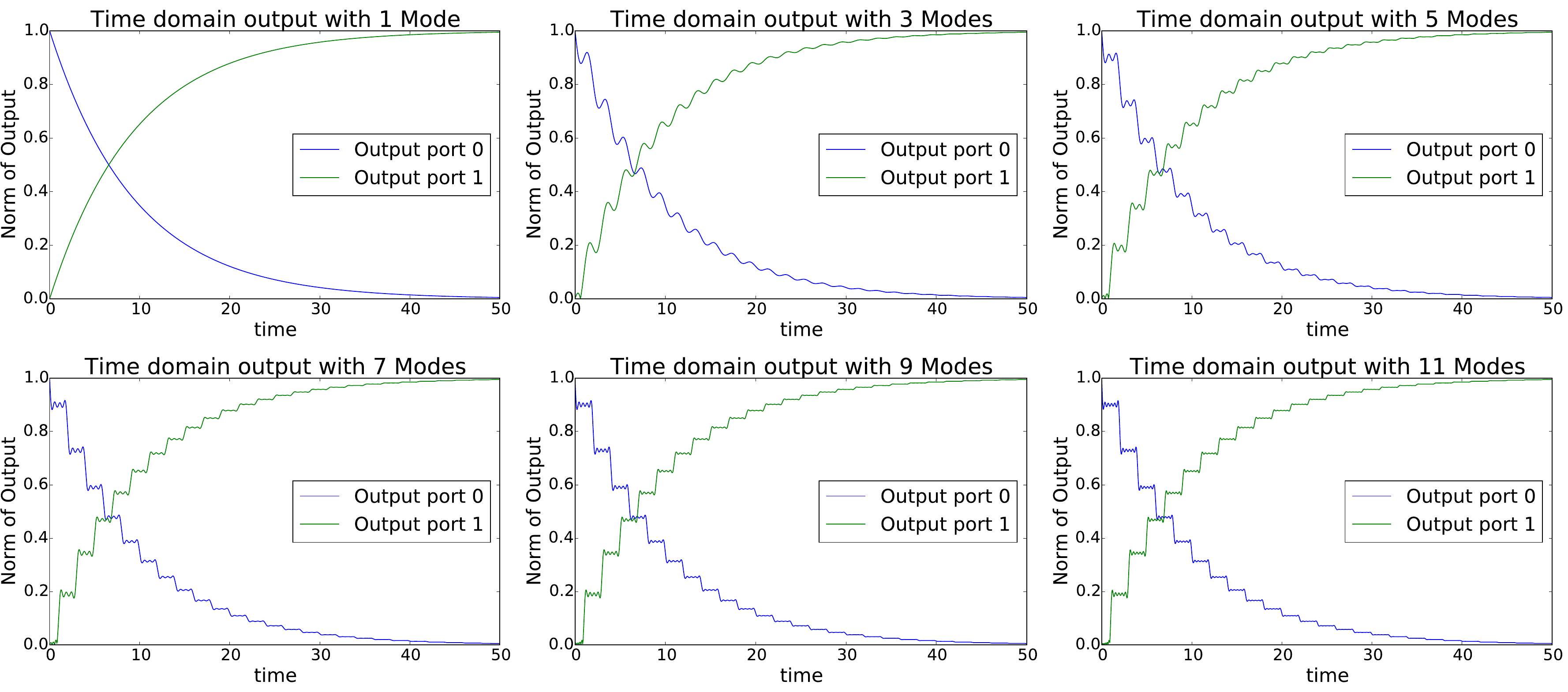}
\caption{\textbf{The output from a Fabry-P\'{e}rot cavity with initial state zero.}   The mirror has reflectivity $r = 0.9$
and the duration of the delay is $1$. The two output ports represent the signal reflected and the signal transmitted through
the cavity. The signal enters the system from port $0$ and exists from both ports $0$ and $1$ (see Figure~\ref{fig:Example_2_time_sketch}).
 In steady-state, the norm of output $0$ converges to zero, an the norm of output $1$ converges to~$1$.
    The stated number of modes in each diagram is the number of modes used in the delay.
    We see that adding more modes eventually converges to a piecewise step function. The steps result from the round-trip
    of the signal inside the cavity.}\label{fig:Example_2_time}
    \end{figure}

\begin{figure}
	\centering
    \includegraphics[width=0.4\textwidth]{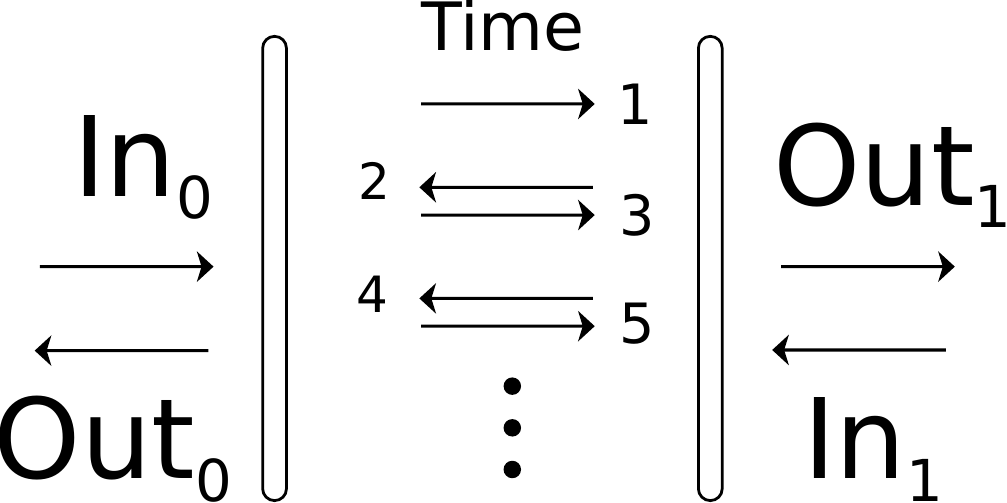}
\caption{\textbf{Interpretation of transient dynamics of the system in the Fabry-P\'{e}rot cavity considered.} Before the system
achieves steady-state, we can learn about the output by tracing the possible paths the signal may follow. The arrows in the
diagram between the two mirrors represent the possible paths the signal may follow before leaving through an output port. We
use the vertical direction to represent time. For each arrow in each output direction, some of the signal leaves through one
of the mirrors. We will therefore see a step in the output as a function of time. This gives a physical interpretation of the steps
seen in Figure~\ref{fig:Example_2_time}.}\label{fig:Example_2_time_sketch}
    \end{figure}

As a simple example, we consider the input-output relationship of a
Fabry-P\'{e}rot cavity with a constant input (\textit{i.e.} $\omega= 0$) at one
of the ports (port $0$) and zero input in the other port (port $1$). In
the steady-state, the signal will be transmitted from the input port
$0$ to the output port $1$. However, if the initial state of the system
is different than the steady-state, we will observe some transient
behavior in the system. This transient behavior is captured by our
simulation and is demonstrated in Figure~\ref{fig:Example_2_time}.
Here, we show the outputs of the two ports based on different numbers
of modes selected to approximate the cavity formed due to the delay. As
the number of modes is increased, we see the signal from the output
ports as a function of time approaches a step function, and we better
reproduce the time-domain dynamics of the network with feedback loops.
The jumps we see in the time-domain correspond physically to times when
a propagating signal arrives at one of the ports. This physical
interpretation is further explained in Figure~\ref{fig:Example_2_time_sketch}.

\section{Conclusion}

In this paper we have utilized the Blaschke-Potapov factorization for
contractive matrix-valued functions to devise a procedure for obtaining
an approximation for the transfer function of physically realizable
passive linear systems consisting of a network of passive components
and time delays. The factorization in our case of interest consists of
two inner functions - the Blaschke-Potapov product, a function of a
particular form having the same zeros and poles as the original
transfer function, and an inner function having no roots or poles (a
singular function). The factors in the Potapov product correspond
physically to resonant modes formed in the system due to feedback,
while the singular term corresponds physically to a feedforward-only
component. We also demonstrate how these two components may be
separated for the type of system considered.

The transfer function resulting from our approximation can be used to
obtain a finite-dimensional state-space representation approximating
the original system for a particular range of frequencies. The
approximated transfer function can also be used to obtain a physically
realizable component in the SLH framework used in quantum optics. Our
approach has the advantage that the zero-pole pairs corresponding to
resonant modes are identified explicitly. In contrast, obtaining a
similar approximation for a feedforward-only component requires
introducing spurious zeros and poles. Our approach has the advantages
that we may retain the numerical values of the zero-pole pairs of the
original transfer function in our approximated transfer function and
that we can conceptually separate these zero-pole pairs from spurious
zeros and poles. These advantages may be important in applications and
extensions of this work.

We hope that in the future our factorization procedure may be extended
to a more general class of linear systems. We also hope to introduce
nonlinear degrees of freedom in a similar way to atoms in the
Jaynes-Cummings model and quantum optomechanical devices in the
presence of modes formed due to optical cavities.

\begin{appendices}

\section{The Cayley Transform}
\label{section:cayley}

Throughout this paper, we work in the frequency domain $\mathbb
{C}$ common in the engineering literature,
where the imaginary axis takes values $i \omega$.
In some of the literature,
a domain related by a bijective conformal transformation is used
instead. The imaginary axis is mapped to the unit circle, and the right
half-plane $\mathbb{C}^+$ is mapped to the unit disk $\mathbb{D}$.
This mapping is known as the Cayley transform, and the two domains are
often called the $s$-domain and $z$-domain, respectively.

The mapping from $\mathbb{C}^+$ to $\mathbb{D}$ is given by
\[
f(z) = \frac{z-1}{z+1}.
\]

The mapping from $\mathbb{D}$ to $\mathbb{C}^+$ is given by
\[
f^{-1} (z)=\frac{1+z}{1-z}.
\]
Sometimes in the literature the upper half-plane is used instead of the
right half-plane, which slightly changes the transformation.

In our notation, a zero $z_k$ of the Blaschke factor in the disc
$
\frac{z_k-z}{1-\overline{z_k} z} \frac{\llvert z_k\rrvert }{z_k}
$
is transformed to a zero in the plane
$\lambda_k = \frac{1+ {z_k} }{1- {z_k} }
$
of the factor
$
e^{i \phi_k} \frac{z-{\lambda_k}}{z + \overline{ \lambda_k}}$,
where $e^{i \phi_k} = \frac{\llvert 1-\lambda_k^2\rrvert }{1-\lambda_k^2}$
contributes only a phase.




%

\section{Unitarity implies a function is inner}
\label{section:max}

(Based on \cite{Dym1997}, Lemma 3 on page 223.)

This section will prove the assertion that
%
\begin{align}
T(z) T^{\dagger}(z)= I \quad \text{for } z \in i \mathbb{R} \quad \implies
\quad T(z) T^{\dagger}(z) \le I \quad \text{for } z \in\mathbb{C}^+,
\end{align}
assuming $T(z)$ is an $N \times N$ matrix-valued analytic (except
possibly at infinity) function bounded in $\mathbb{C}^+$.


For any appropriately sized unit vectors ${u}$, ${v}$, we have from the
Cauchy-Schwartz inequality and the maximum modulus principle that
%
\begin{align}
\bigl\llvert \bigl\anglel u,T(z) v \bigr\angler \bigr\rrvert ^2 \le
\llVert u\rrVert \bigl\llVert T(z) v \bigr\rrVert \le1.
\end{align}
Since the $u$, $v$ were arbitrary, the assertion follows.

\section{Potapov factorization theorem and non-passive linear systems}
\label{section:active}

In this section we cite the Potapov factorization theorem for
$J$-contractive matrices.
This factorization, when applicable, consists of
several terms which each have a different property.
For this paper we will only need a special case for the theorem,
but we cite the full version because it may be conductive to extensions
of the work in this paper.
In particular, a related frequency-domain condition of physical
realizability is discussed in \cite{Petersen2010} and \cite{Shaiju2012}.

\subsection{Definitions}

\label{section:defs}

First, we introduce some terminology common in the literature.
Let $D = \mathbb{D}$ or $\mathbb{C}_+$ (the unit disc or right
half-plane, respectively).
Further let $J$ be the signature matrix
%
\begin{align}
J = %
\begin{pmatrix}
I_m & 0 \\
0 & - I_r
\end{pmatrix} %
,
\end{align}
for some $m$, $r$.
A matrix-valued function $M(z)$ is called:
\begin{enumerate}
\item$J$-contractive, when
$M(z)JM(z)^{\dagger}\le J\rq{}$ for z in ${D}$,
\item$J$-unitary when
$M(z)JM(z)^{\dagger}= J\rq{}$ for $z$ on $\partial D$,
\item$J$-inner (or $J$-lossless) when $M(z)$ is $J$-unitary and
$J$-contractive.
\end{enumerate}
When $J = I$, we drop the $J$ in the definition.

\emph{Definition.} [(Stieltjes multiplicative integral (from \cite{Potapov1955}))]
Let $K(t)$ (with $a \le t \le b$) be a monotonically increasing
family of $J$-Hermitian matrices with $t = \text{tr} [K(t) J] $ and
$f(t)$ (${a \le t \le b}$) be a continuous scalar function. Then the
following limit exists
%
\begin{align}
\label{equ:mult_int_def} \lim_{\max\Delta t_j \to0 } e^{f(\theta_0) \Delta K(t_0)} e^{f(\theta_1) \Delta K(t_1)}\cdots
e^{f(\theta_{n-1}) \Delta K(t_{n-1})},
\end{align}
where we take $a = t_0 \le\theta_0 \le t_1 \le\cdots \le t_n = b$.
The limit is denoted
\[
\overset{\curvearrowright} {
\int_a}^b e^{f(t)\,dK(t)}
\]
and is called the multiplicative integral.


 \subsection{Potapov Factorization}
 
 \label{section:potapov}
 
 
In much of the mathematical literature the domain of the transfer
function is transformed via the Cayley transform
(see Appendix~\ref{section:cayley}).
This changes how some of the terms in the factorization are written,
but not the fundamental features of the factorization.
%

Next we will cite some of the theorems by Potapov.
In the case $J=I$ the function $T$ satisfies the unitarity
condition $T(z)T^{\dagger}(z) = I$ on the boundary of the disc or
half-plane (depending on the domain taken).


The fundamental factorization theorem by Potapov characterizes the
class of $J$-contractive matrices.

\emph{Theorem.} [(Adapted from \cite{Potapov1955})]
Let $T(z)$ be a [meromorphic] $J$-contractive matrix function
in the unit circle $\llvert z\rrvert  < 1$, and suppose that $\det( T(z))$
does not vanish identically; then we can write
%
\begin{align}
\label{equ:omega_potapov} T(z) = B_\infty(z) B_0(z) \overset{
\curvearrowright} {
\int_0^\ell} \exp \biggl( \frac{z+e^{i\theta(t)}}{z-e^{i\theta(t)}}
\,dK(t) \biggr).
\end{align}
Here
%
\begin{align}
B_\infty(z) = \prod_k U_k
\begin{pmatrix}
I & 0 \\
0 & \frac{1 - \overline\mu_k z}{\mu_k - z}\frac{\mu_k}{\llvert \mu_k\rrvert
}I_{q_k\rq{}}
\end{pmatrix} %
U_k^{-1}
\end{align}
is a product of elementary factors associated with the poles $\mu_k$ of
the matrix function $T(z)$ inside the unit circle, $q_k\rq{} \le
q$, and $U_k$ is a J-unitary matrix;
%
\begin{align}
B_0 (z) = \prod_k V_k
\begin{pmatrix}
\frac{\lambda_k - z}{1 - \overline\lambda_k z} \frac{| \lambda
_k | }{\lambda_k} I_{p_k\rq{} } & 0 \\
0 & I
\end{pmatrix} %
V_k^{-1}
\end{align}
is a product of elementary factors associated with the zeros in $\llvert z\rrvert
<1$ of the determinant of the matrix function $T_\infty(z) =
B_\infty^{-1} (z) T(z)$, which is holomorphic in $\llvert z\rrvert  < 1$, $p_j\rq{}
\le p$, $V_j$ is a $J$-unitary matrix;
the last term is the Stieltjes integral, where $K(t) J$ is a
monotone increasing family of Hermitian matrices such that $t = \tr
[K(t) J] = t$. Here $\theta\in[0,2 \pi]$ is a monotonically
increasing function

The integral in the above expression is known as the Riesz-Herglotz
integral. It captures the effects of the zeros and poles that occur on
the boundary as well as effects not due to zeros or poles.
\emph{Theorem.} (from~\cite{Potapov1955}):
An entire matrix function $T(z)$, $J$-contractive in the
right half-plane and $J$-unitary on the real axis, can be represented
in the form
%
\begin{align}
\label{equ:singular_factorization} T(z) = T(0) \overset{\curvearrowright} {
\int_0^\ell} e^{-z K(t)\,dt},
\end{align}
where $K(t)$ is a summable non-negative definite $J$-Hermitian matrix,
satisfying the condition $\tr [K(t) J ] = 1$.

\section{Using the Pad\'{e} approximation for a delay}
\label{section:pade_approximation}
For a system involving only feedforward (\textit{i.e.} no signal ever feeds
back), no poles or zeros will be found in the transfer function. For
this reason, the zero-pole interpolation cannot naturally reproduce a
transfer function to approximate the system. Instead, a different
approach is needed to obtain an approximation for the state-space
representation for systems of this kind. Still, any finite-dimensional
state-space representation will have poles and zeros in its transfer
function. If we use such a system as an approximation of a delay, these
zeros and poles will be spurious but unavoidable.

The Pad\'{e} approximation is often used to approximate delays in
classical control theory \cite{Lam1993}. When using the $[n,n]$
diagonal version of the Pad\'{e} approximation to obtain a rational
function approximating an exponential, we obtain
%
\begin{align}
e^{-Tz} \approx\exp_{[n,n]}(-Tz) 
=
\frac{Q_n(zT)}{Q_n(-zT)}.
\end{align}
The $Q_n(z)$ is a polynomial of degree $n$ with real coefficients.
Because of this, its roots come in conjugate pairs. As a result, we can
write the Pad\'{e} approximation as a product of Blachke factors:
%
\begin{align}
\exp_{[n,n]}(-Tz) = \prod_n -
\frac{z + \overline{{p}_n }}{z+ p_n}.
\end{align}
In particular, note that the approximation preserves the unitarity
condition, and is therefore physically realizable.
For this reason, it is possible to approximate time delays with this
approximation.

\begin{figure}
    \centering
    \includegraphics[width=1.0\textwidth]{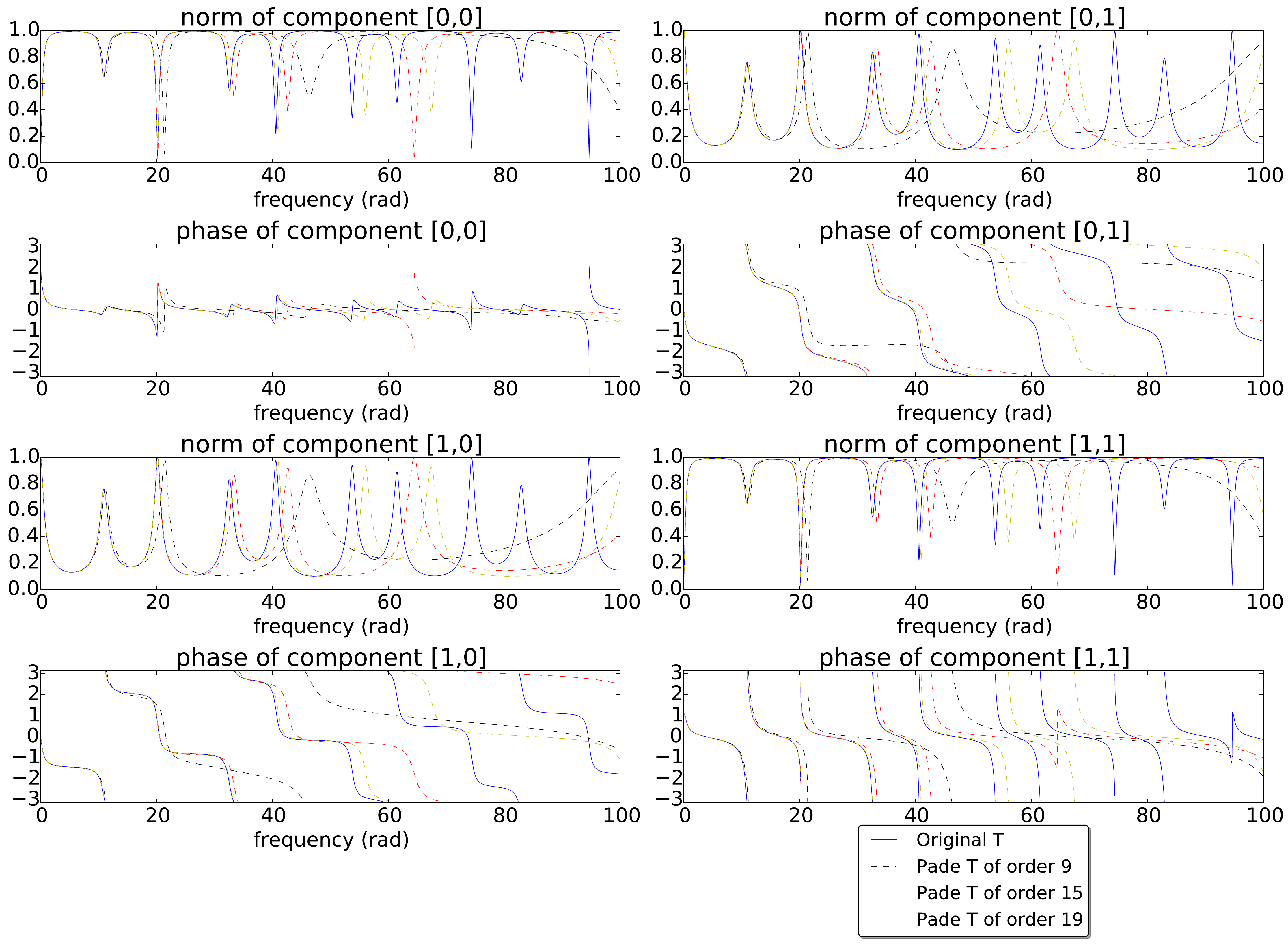}
\caption{\textbf{The approximated transfer functions resulting from applying the Pad\'{e}  approximation on individual
delays in the network. } Although we see that this approximation converges, the peaks often do not occur at the correct locations,
as opposed to the approximations resulting from the zero-pole
interpolation (compare to  Figure~\protect\ref{fig:Example_4_graph}).}\label{fig:Pade_example}
    \end{figure}

Although this approach may be useful for the case of feedforward-only delays,
in the case of delays with feedback this may produce
undesirable results.
To illustrate this, we introduce an example where the Pad\'{e} approximation
is used in order to produce an approximated transfer function for the network
discussed in Section~\ref{sec:zero_pole_success}.
For this approximation, the order used for the approximation of
each delay is chosen to be roughly proportional to the duration of the delay.
Figure~\ref{fig:Pade_example} illustrates the approximated transfer
function. Please compare this result to Figure~\ref{fig:Example_4_graph}, where we have used the zero-pole interpolation.

We see in Figure~\ref{fig:Pade_example} that the peaks in the
approximating functions often do not
occur in same locations as the peaks of the original transfer function.
This may be problematic when attempting to simulate many physical
systems for which the locations of the peaks correspond to particular
resonant
frequencies that have physical relevance.
For instance, if one chooses to introduce other components to the system,
such as atoms,
the resonant frequencies due to the trapped modes of the network
must be
described accurately or else the resulting dynamics
of the approximating system may not correspond to the
true dynamics of the physical system.
For this reason, using the Pad\'{e} approximation may not always be
the best choice.

\section{Blaschke-Potapov Product in the limit ${\Re(z) \to -\infty}$}
\label{section:limit_argument}

In Section~\ref{section:singular}, we proposed a criterion for checking
when the multiplicative integral component in Section~\ref{section:factorization_theorem} was not necessary. We assumed that the
Blaschke-Potapov product converged to a nonzero constant in the limit
${\Re(z) \to-\infty}$.
In this section, for demonstrative purposes we show this is the case
for the example of a single trapped cavity discussed in Section~\ref{section:first_example}.


In order to examine the convergence of the product in Eq. (\ref
{equ:prod_ex}) of Section~\ref{section:first_example}, we write it in
the following way.
We observe by taking the logarithm with an appropriate branch cut that
%
\begin{align}
\label{equ:prod_as} \prod_n \bigl(1+
a_n(z)\bigr)
\end{align}
converges if and only if the infinite sum
%
\begin{align}
\label{equ:log_expansion} \sum_n a_n(z)
\end{align}
converges, assuming $\sum_n \llvert a_n(z)\rrvert ^2 $ converges. We take
%
\begin{align}
a_n(z) = 1 - \frac{z+p_n }{z-\overline{p_n} } = \frac{2 \Re(p_n)}{z - p_n}.
\end{align}
Using the relation
%
\begin{align}
\cot(z) = \sum_{n \in\mathbb{Z}} \frac{1}{z - n \pi},
\end{align}
we get that
%
\begin{align}
\lim_{\Re(z) \to-\infty} \sum_n
a_n(z) =-\ln(r).
\end{align}
Actually, higher-order terms of the logarithm expansion of Eq. (\ref
{equ:prod_as}) go to zero in this limit, so we get that
%
\begin{align}
\lim_{\Re(z) \to-\infty}\tilde T(z) = \lim_{\Re(z) \to-\infty}-\prod
_{n \in\mathbb{Z}} \biggl( \frac
{z+\overline{p_n} }{z-p_n } \biggr)= \lim
_{\Re(z) \to-\infty} - \prod_n \bigl(1+
a_n(z)\bigr) = -1/r.
\end{align}
This shows the desired result that the infinite product in this limit
goes to a nonzero constant. Also interestingly, we have been able to
compute this value and remark that it is indeed equal to $\lim_{\Re(z)
\to-\infty} T(z)$.

The above example with a single trapped cavity formed is illustrative
of typical behavior for more complicated systems formed by networks of
beamsplitters and time delays. The zero-pole pairs of the system occur
in a region of bounded positive real part, and roughly uniformly along
the imaginary axis. This suggests that the Blaschke-Potapov product
resulting from the zero-pole interpolation will converge to a nonzero
constant in the ${\Re(z) \to-\infty}$ limit under quite general circumstances.\end{appendices}

\section*{Acknowledgements:}

This work has been supported by ARO under grant W911-NF-13-1-0064. Gil
Tabak was supported by the Department of Defense (DoD) through the
National Defense Science \& Engineering Graduate Fellowship (NDSEG)
Program and by a Stanford University Graduate Fellowship (SGF). We
would also like to thank Ryan Hamerly, Nikolas Tezak, and David Ding
for useful discussions while this work was being prepared.

%
%
%
%

\bibliographystyle{plain}

\bibliography{paperBibliography}

\end{document}